\documentclass[aps,prd,onecolumn,showpacs,showkeys,amsmath,amssymb,nofootinbib]{revtex4}
\usepackage{graphicx}
\usepackage{dcolumn}
\usepackage{amsmath}
\usepackage{bm}
\usepackage{yhmath}
\usepackage{hyperref}
\usepackage{subfigure}
\usepackage{multirow}

\begin{document}

\def\Tr{{\rm Tr}}
\newcommand{\sla}{\not\!}
\renewcommand\arraystretch{1.8}
\pagenumbering{arabic} \allowdisplaybreaks

\title{Chiral Perturbation Theory and the $\bar B \bar B$ Strong Interaction}

\author{Zhan-Wei Liu}\email{liuzhanwei@pku.edu.cn}
\author{Ning Li}\email{leening@pku.edu.cn}
\author{Shi-Lin Zhu}\email{zhusl@pku.edu.cn}
\affiliation{Department of Physics
and State Key Laboratory of Nuclear Physics and Technology\\
Peking University, Beijing 100871, China }

\begin{abstract}

We have calculated the potentials of the heavy (charmed or
bottomed) pseudoscalar mesons up to $O(\epsilon^2)$ with the heavy
meson chiral perturbation theory. We take into account the
contributions from the football, triangle, box, and crossed
diagrams with the 2$\phi$ exchange and one-loop corrections to the
contact terms. We notice that the total 2$\phi$-exchange potential
alone is attractive in the small momentum region in the channel
${\bar B \bar B}^{I=1}$, ${\bar B_s \bar B_s}^{I=0}$, or ${\bar B
\bar B_s}^{I=1/2}$, while repulsive in the channel ${\bar B \bar
B}^{I=0}$. Hopefully the analytical chiral structures of the
potentials may be useful in the extrapolation of the heavy meson
interaction from lattice QCD simulation.

\end{abstract}



\maketitle

\section{Introduction}\label{secIntr}

Since the discovery of $X(3872)$~\cite{Choi2003}, many
charmonium-like and bottomonium-like states such as
$X(3940)$~\cite{Abe2007}, $X(4160)$~\cite{Pakhlov2008}, {\it et
al.} have been observed in the past decade. The charmonium-like
state $X(3872)$ was first observed by the Belle Collaboration in
the exclusive decay process $B^{\pm}\to K^{\pm}\pi^+\pi^-J/\psi$.
Last year the Belle Collaboration observed two charged
bottomonium-like resonances $Z_b(10610)$ and $Z_b(10650)$ in the
hidden-bottom decay channels $\pi^{\pm}\Upsilon(nS)$ ($n=1,~2,~3$)
and $\pi^{\pm}h_b(mP)$ ($m=1,~2$) of
$\Upsilon(5S)$~\cite{Adachi2011}.

Some of these new states including the above two charged $Z_b$
states do not fit into the conventional quark model framework.
Various theoretical approaches including the lattice
QCD~\cite{Yang:2012my}, the QCD sum rule \cite{Chen2011}, and
the quark model~\cite{Liu:2008qb} have been employed to interpret
the underlying structure of these new states. Despite huge
experimental and theoretical efforts, the nature of some of these
exotic states is still elusive.

For example, the interpretation of $X(3872)$ remains challenging
since the discovery in
2003. One popular
speculation is that $X(3872)$ is a molecular state composed of a
pair of heavy mesons \cite{AlFiky:2005jd,Thomas:2008ja,Liu2009c}. Similarly, the two charged
$Z_b(10610)$ and $Z_b(10650)$ states are proposed as the $B\bar{B}^*$ and
$B^*\bar{B}^*$ molecule states within the one boson exchange (OBE)
framework ~\cite{Sun:2011uh,Ohkoda:2011vj}.

Besides the charmonium-like and bottomonium-like states, the
possible existence of some molecular candidates composed of $\bar
B \bar B$ mesons and $DD$ mesons is also very interesting. If the
attractive interaction is strong enough between the heavy meson
pair, this kind of states may exist. Their behavior will be very
similar to the deuteron which is composed of two nucleons. There
have been some investigation of these interesting states within
the OBE model.

However, the interaction potential derived from the OBE model
contains several phenomenological coupling constants and cutoff
parameters, which should in principle be extracted through fitting
to the experimental data. Unfortunately, there is not much
experimental information on the strong interaction between the
light meson and heavy meson. It will be very desirable to derive
the strong interaction between the heavy meson pair with a
model-independent approach. Especially many new states such as
$X(3872)$ and the two $Z_b$ states lie very close to the threshold.
Within these very loosely bound systems, the long-range pion
exchange force should play an important role. Therefore the chiral
perturbation theory provides a natural framework to investigate
the heavy meson strong interaction. In this work, we shall derive
the heavy pseudoscalar meson potential order by order.

Chiral perturbation theory ($\chi$PT) is a model-independent tool
to study the chiral dynamics of heavy hadrons. Heavy hadron
$\chi$PT is frequently used for the system made up with a single
heavy hadron and light pseudoscalar mesons because of its explicit
power counting \cite{Cho1993,Bernard1997,Hemmert1998,Liu2012b,Geng2008}. The scattering matrix
can be expanded order by order in the small parameter
$\epsilon=p/\Lambda_\chi$, where $p$ represents either the
momentum of the light pseudoscalar mesons or the residual momentum
of the heavy hadrons in the nonrelativistic limit, while
$\Lambda_\chi$ represents either the scale of chiral symmetry
breaking or the mass of heavy hadrons. The power counting
guarantees that one can just calculate some limited Feynman
diagrams and obtain the scattering matrix at the certain order.

Weinberg developed a new formalism and first extended the chiral
perturbation theory to the two nucleon system
\cite{Weinberg1990,Weinberg1991}. Since his pioneering work, the
modern nuclear force has been built upon the chiral effective
field theory
\cite{Bernard1995,Ordonez1996,Kaiser1997758,Kaiser1998,Epelbaoum1998,Epelbaoum2000,Machleidt2011,Kaiser2001a,Epelbaum2003,Epelbaum2005,Zhu2005}.
Such a formalism was used extensively to investigate the various
few-nucleon observables such as the partial-wave analysis,
few-nucleon scattering, and reaction.

\begin{figure}[!htbp]
\centering
\scalebox{0.8}{\includegraphics{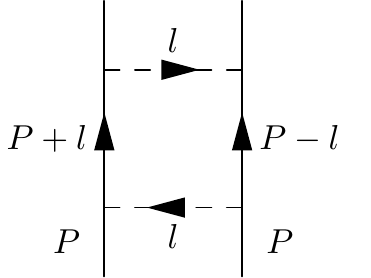}}\\
\caption{The box diagram. The solid line represents the nucleon,
and the dashed line represents the pion.
 } \label{fig2PR}
 \end{figure}

As pointed out by Weinberg \cite{Weinberg1990,Weinberg1991}, the
power counting of the two nucleon scattering matrix is broken by
the double poles of the heavy hadrons in some 2-particle-reducible
(2PR) Feynman diagrams. Let us illustrate this point with the box
diagram in Fig. \ref{fig2PR}. The Feynman amplitude can be written
as
\begin{equation}
  i\int d^4 l \frac{i}{l^0+P^0+i\varepsilon}\frac{i}{-l^0+P^0+i\varepsilon}\times \cdots
  =i\int d l^0 \frac{i}{l^0+P^0+i\varepsilon}\frac{i}{-l^0+P^0+i\varepsilon}\int d^3 l \cdots,
\end{equation}
where we omit the parts relevant to the pion which preserves the
power counting. We will focus the integral with $l^0$, and work it
out by closing the $l^0$ contour integral in the lower half-plane
\begin{equation}
  \mathcal I\equiv i\int d l^0 \frac{i}{l^0+P^0+i\varepsilon}\frac{i}{-l^0+P^0+i\varepsilon}=\frac{\pi}{P^0+i\varepsilon}
  \approx \frac{\pi}{\vec P^2/(2m_N)+i\varepsilon} .  \label{eq2PR}
\end{equation}
The naive power counting predicts that $\mathcal I$ should be
$O(1/|\vec P|)$. But it is $O(m_N/|\vec P|^2)$ from Eq.
(\ref{eq2PR})! $\mathcal I$ is actually enhanced by a large factor
$m_N/|\vec P|$ compared to the naive power counting prediction.

With Weinberg's formalism, we do not directly calculate the
scattering matrix of the few hadrons perturbatively with the heavy
hadron $\chi$PT due to the 2PR diagrams. Instead, we focus on the
potential. In the derivation of the hadron-hadron potential, one
takes into account the 2-particle-irreducible (2PI) parts of the
Feynman diagrams only and calculate the potential of few hadrons
perturbatively with the correct power counting. Afterwards, one
can obtain the scattering matrix with the potential by solving the
nonperturbative equations such as Schr{\"o}dinger equations,
Lippmann-Schwinger equations, and so on. The 2PR contributions
will be recovered when solving the nonperturbative equations.

The reliable hadron-hadron potential is a necessary input for
getting the scattering amplitude or phase shift of the hadrons. It
is also essential to explore the existence of the heavy hadron
molecules. For example, the binding energy or size of the
molecular states can be obtained from the potentials of the
hadrons by solving the Schr{\"o}dinger or Lippmann-Schwinger
equations.

In this work, we shall use Weinberg's formalism to derive the
$\bar B \bar B$ potentials in four independent channels up to
1-loop level with heavy meson chiral perturbation theory
(HM$\chi$PT). We include the heavy vector $\bar B^*$ mesons as
explicit degrees besides the $\bar B$ and light pseudoscalar
$\phi$ mesons since the $\bar B$ and $\bar B^*$ mesons would form
a degenerate doublet in the limit of heavy quark symmetry. We
count the mass difference $\Delta$ between $\bar B$ and $\bar B^*$
mesons as $O(\epsilon^1)$. The potentials of the $\bar B\bar B$
mesons start at $O(\epsilon^0)$. We will investigate the
corrections up to $O(\epsilon^2)$.

This paper is organized as follows. In Sec. \ref{secLagr}, we list
the Lagrangians of HM$\chi$PT. In Sec. \ref{secPtntl}, we present
the expressions of the $\bar B\bar B$ potential, which include the
tree-level diagram contributions at the leading order and the loop
corrections at $O(\epsilon^2)$. In Sec. \ref{secNum}, we give the
numerical results of the $\bar B\bar B$ potentials in the first
subsection. Then we present the results of the potentials of the
$D D$ mesons in the second
subsection. We compare the results within different
schemes in Sec. \ref{secCom}. Sec. \ref{secSum} is a short summary.

\section{Lagrangians with HM$\chi$PT}\label{secLagr}

The leading order $\bar B\bar B$ potential is at $O(\epsilon^0)$ and
receives only the contribution from the tree-level diagrams made
up of the vertices of the leading order Lagrangians in Eq.
(\ref{HHHH}). The corrections start at $O(\epsilon^2)$. They
contain the contributions of the 1-loop diagrams generated by the
leading Lagrangians and the contributions of tree diagrams
generated by the Lagrangians at higher order.

The leading Lagrangians are
\begin{eqnarray}
\mathcal L^{(0)}_{4H}&=&D_{a}\Tr[H \gamma_\mu \bar H ]\Tr[ H
\gamma^\mu\bar H] +D_{b}\Tr[H \gamma_\mu\gamma_5 \bar H ]\Tr[ H
\gamma^\mu\gamma_5\bar H] \nonumber\\&& +E_{a} \Tr[H
\gamma_\mu\lambda^a \bar H ]\Tr[ H \gamma^\mu\lambda_a\bar H]
+E_{b}\Tr[H \gamma_\mu\gamma_5\lambda^a \bar H ]\Tr[ H
\gamma^\mu\gamma_5\lambda_a\bar H],  \label{HHHH}
\\
\mathcal L^{(1)}_{H\phi}&=&-\langle (i v\cdot \partial H)\bar H
\rangle
                         +\langle H v\cdot \Gamma \bar H \rangle
                         +g\langle H \sla u \gamma_5 \bar H\rangle
                         -\frac18 \Delta \langle H \sigma^{\mu\nu} \bar H \sigma_{\mu\nu} \rangle,\label{L1}
\end{eqnarray}
where the number in the round bracket represents the chiral
dimension, $v_\mu = (1, \vec 0)$ is the velocity of a slowly
moving heavy meson, and $H$ represents the $\bar B$ and $\bar B^*$
doublet in the heavy quark symmetry limit,
\begin{eqnarray}
 && H=\frac{1+\sla v}{2}\left(P^*_\mu\gamma^\mu+iP\gamma_5\right),\quad
 \bar H=\gamma^0 H^\dag \gamma^0 = \left(P^{*\dag}_\mu \gamma^\mu+iP^\dag \gamma_5\right) \frac{1+\sla v}{2},\\
 && P=(B^-, \bar B^0,\bar B^0_s), \quad P^*_\mu=(B^{*-}, \bar B^{*0}, \bar B_s^{*0})_\mu. \label{PPs}
\end{eqnarray}
The pseudoscalar meson field, chiral connection and axial vector
field are defined as follows,
\begin{equation}
\Gamma_\mu = {i\over 2} [\xi^\dagger, \partial_\mu\xi],\quad
u_\mu={i\over 2} \{\xi^\dagger, \partial_\mu \xi\},\quad \xi =
\exp(i \phi/2f),
\end{equation}
\begin{equation}
 \phi=\sqrt2\left(
\begin{array}{ccc}
\frac{\pi^0}{\sqrt2}+\frac{\eta}{\sqrt6}&\pi^+&K^+\\
\pi^-&-\frac{\pi^0}{\sqrt2}+\frac{\eta}{\sqrt6}&K^0\\
K^-&\overline{K}^0&-\frac{2}{\sqrt6}\eta
\end{array}\right).
\end{equation}

The Lagrangian $\mathcal L^{(0)}_{4H}$ generates the contact
interaction terms of the four bottomed mesons while $\mathcal
L^{(1)}_{H\phi}$ depicts the interaction between the heavy mesons
and light pseudoscalar mesons. The other contact terms with
different Lorentz structures at the leading order are not
independent. Actually they are linear combinations of terms in
$\mathcal L^{(0)}_{4H}$, so we do not need them. For example, the
term $\Tr[H \gamma_\mu \bar H H \gamma^\mu\bar H]$ can be
expressed as the linear combination of terms in $\mathcal
L^{(0)}_{4H}$ by Fierz transformation. In the heavy meson limit,
$\Tr[H \bar H ]\Tr[ H \bar H]$ and $\Tr[H \sigma_{\mu\nu} \bar H
]\Tr[ H \sigma^{\mu\nu}\bar H]$ can be absorbed by readjusting the
coefficients $D_a$ and $D_b$ of $\mathcal L^{(0)}_{4H}$,
respectively. The term $\Tr[H \gamma_5 \bar H ]\Tr[ H \gamma^5
\bar H]$ vanishes in the heavy meson limit. Similar conclusions
hold for the terms containing $\lambda^a$ such as $\Tr[H
\gamma_5\lambda^a \bar H ]\Tr[ H \gamma^5 \lambda_a\bar H]$ etc.

The ranges of the couplings $D_a$, $D_b$, $E_a$, and $E_b$ were
estimated in the $D\bar D$ case by fixing the masses of $X(3872)$,
$X(3915)$, and $Y(4140)$ and the isospin breaking branching ratio
of $X(3872)$ in Ref. \cite{Hidalgo2012}. Their values lie from
several to tens of ${\rm GeV^{-2}}$ with positive signs.

The $O(\epsilon^2)$ Lagrangian $\mathcal L^{(2)}_{4H}$ will also
contribute to the potentials, which read
\begin{eqnarray}
\mathcal L^{(2,h)}_{4H}&=& D_{a}^h\Tr[H \gamma_\mu \bar H ]\Tr[ H
\gamma^\mu\bar H] \Tr(\chi_+) +D_{b}^h\Tr[H \gamma_\mu\gamma_5
\bar H ]\Tr[ H \gamma^\mu\gamma_5\bar H]\Tr(\chi_+) \nonumber\\&&
+E_{a}^h \Tr[H \gamma_\mu\lambda^a \bar H ]\Tr[ H
\gamma^\mu\lambda_a\bar H]\Tr(\chi_+) +E_{b}^h \Tr[H
\gamma_\mu\gamma_5\lambda^a \bar H ]\Tr[ H
\gamma^\mu\gamma_5\lambda_a\bar H]\Tr(\chi_+),  \label{eqL2h}
\\
\mathcal L^{(2,d)}_{4H}&=& D_{a}^d\Tr[H \gamma_\mu\tilde\chi_+
\bar H ]\Tr[ H \gamma^\mu\bar H] +D_{b}^d\Tr[H
\gamma_\mu\gamma_5\tilde\chi_+ \bar H ]\Tr[ H
\gamma^\mu\gamma_5\bar H] \nonumber\\&& +E_{a}^d d^{abc} \Tr[H
\gamma_\mu\lambda_a \bar H ]\Tr[H \gamma^\mu\lambda_b \bar H
]\Tr[\tilde\chi_+\lambda_c ] +E_{b}^d  d^{abc} \Tr[H
\gamma_\mu\gamma_5\lambda_a \bar H ]\Tr[H
\gamma^\mu\gamma_5\lambda_b \bar H ]\Tr[\tilde\chi_+\lambda_c ], \label{eqL2d}
\\
\mathcal L^{(2,v)}_{4H}&=& \{D_{a1}^v\Tr[(v\cdot D H) \gamma_\mu
(v\cdot D \bar H) ]\Tr[ H \gamma^\mu\bar H] +D_{a2}^v\Tr[(v\cdot D
H) \gamma_\mu \bar H ]\Tr[ (v\cdot D H) \gamma^\mu\bar H]
\nonumber\\&& +D_{a3}^v\Tr[(v\cdot D H) \gamma_\mu \bar H ]\Tr[  H
\gamma^\mu(v\cdot D \bar H)] +D_{a4}^v\Tr[((v\cdot D)^2 H)
\gamma_\mu \bar H ]\Tr[  H \gamma^\mu\bar H ] \nonumber\\&&
+D_{b1}^v\Tr[(v\cdot D H) \gamma_\mu\gamma_5 (v\cdot D \bar H)
]\Tr[ H \gamma^\mu\gamma_5\bar H]+... +E_{a1}^v \Tr[(v\cdot D H)
\gamma_\mu\lambda^a (v\cdot D \bar H) ]\Tr[ H
\gamma^\mu\lambda_a\bar H]+... \nonumber\\&& +E_{b1}^v\Tr[(v\cdot
D H) \gamma_\mu\gamma_5\lambda^a (v\cdot D \bar H) ]\Tr[ H
\gamma^\mu\gamma_5\lambda_a \bar H]+...\} +\text{H.c.},   \label{eqL2v}
\\
\mathcal L^{(2,q)}_{4H}&=& \{D_1^q\Tr[(D^\mu H) \gamma_\mu\gamma_5
(D^\nu \bar H) ]\Tr[ H \gamma_\nu\gamma_5\bar H] +D_2^q\Tr[(D^\mu
H) \gamma_\mu\gamma_5 \bar H ]\Tr[ (D^\nu H)
\gamma_\nu\gamma_5\bar H] \nonumber\\&& +D_3^q\Tr[(D^\mu H)
\gamma_\mu\gamma_5 \bar H ]\Tr[ H \gamma_\nu\gamma_5(D^\nu \bar
H)] +D_4^q\Tr[(D^\mu D^\nu H) \gamma_\mu\gamma_5 \bar H ]\Tr[ H
\gamma_\nu\gamma_5 \bar H] \nonumber\\&& +E_1^q\Tr[(D^\mu H)
\gamma_\mu\gamma_5 \lambda^a(D^\nu \bar H) ]\Tr[ H
\gamma_\nu\gamma_5\lambda_a\bar H] +...\}+\text{H.c.},\label{eqL2q}\\
\cdots&& \nonumber
\end{eqnarray}
where $d^{abc}$ is the totally symmetric structure constant of
SU(3) group, and
\begin{equation}
\tilde\chi_\pm=\chi_\pm-\frac13\Tr[\chi_\pm],\quad
\chi_\pm=\xi^\dagger\chi\xi^\dagger\pm\xi\chi\xi,\quad
\chi=\mathrm{diag}(m_\pi^2,\, m_\pi^2,\, 2m_K^2-m_\pi^2).
\end{equation}

The low-energy constants (LECs) in Eqs. (\ref{eqL2h})-(\ref{eqL2q}) contain both
the infinite and finite parts. We will use the infinite parts to
cancel the divergence introduced by the loop diagrams. We are
unable to determine the finite parts of the
LECs due to lack of experimental data right now, which we tend
to neglect for the moment. However, these LECs at $O(\epsilon^2)$
should be included in a complete analysis in the future when more
experimental data are available.

There will be devoted efforts to study the new resonances composed
of a pair of $\bar B^{(*)}B^{(*)}$ at the approved SuperBelle,
KEK.
The investigation of these systems may reveal
the interaction between $\bar B^{(*)}B^{(*)}$. For example, one
may know whether the interaction is attractive at some distance.

Moreover, right now there exists dedicated huge efforts to study
the $\bar D^{(*)} D^{(*)}$ interaction through the decays of the
excited charmonium resonances at BESIII/BEPCII at IHEP, Beijing.
In quantum field theory, the $\bar B^{(*)} B^{(*)}$ interaction
could be related to the $\bar B^{(*)}\bar B^{(*)}$ except the short distance
part due to the annihilation in the $\bar B^{(*)} B^{(*)}$
channel \cite{Timmers1984, Hippchen1991}.

Besides, using lattice QCD within the L\"uscher's formalism, the
scattering length and the scattering phase shifts have been
studied for pion-pion scattering, $D^*\bar D_1$ scattering, and so
on \cite{Luscher1990, Gupta1993, Ming2010}. If there were lattice
studies about the $\bar B \bar B$ or $DD$ scattering in different
partial waves and different channels, we could fix some parameters
or reduce the number of independent parameters with the lattice
information.
In Appendix \ref{secFit} we fit some LECs with the results of quenched lattice QCD.
In our subsequent work, we also plan to reduce the
number of independent LECs by assuming large $N_c$ and heavy quark
symmetry as used in Ref. \cite{Lutz2011}.

$\mathcal L^{(2,h)}_{4H}$ and $\mathcal L^{(2,d)}_{4H}$ are made
up of four heavy meson fields, $\Tr(\chi_+)$ and traceless
$\tilde \chi_+$. The LECs in $\mathcal L^{(2,h)}_{4H}$ and
$\mathcal L^{(2,d)}_{4H}$ will absorb the divergent parts from the
one-loop diagrams which are proportional to $m_\phi^2$. $\mathcal
L^{(2,v)}_{4H}$ will absorb the divergent parts proportional to
the square of the external line energy. There are also divergent
parts proportional to the mass difference $\Delta$, which will
vanish in the heavy meson symmetry limit. These divergences can be
absorbed by the additional four heavy meson interaction terms
proportional to $\Delta$. $\mathcal L^{(2,q)}_{4H}$ does not
contribute to the renormalization of the $\bar B\bar B$ potentials. Instead
it will contribute to the $\bar B\bar B^*$ and $\bar B^*\bar B^*$ potentials.

\section{Potentials with HM$\chi$PT}\label{secPtntl}

With the strict isospin symmetry, there are only four independent
potentials for the channels ${\bar B \bar B}^{1}$, ${\bar B \bar
B}^{0}$, ${\bar B_s \bar B_s}^{0}$, and ${\bar B \bar B_s}^{1/2}$.
The superscript represents the isospin of the channel. At the
leading order, the potentials of the bottomed mesons only receive
the contributions from the tree diagrams with the contact terms in
$\mathcal L^{(0)}_{4H}$,
\begin{eqnarray}
V_{{\bar B \bar B}^{1}}^{(0)}=-2 D_{a}-\frac{8}{3} E_{a}, \quad
V_{{\bar B \bar B}^{0}}^{(0)}=0, \quad V_{{\bar B_s \bar
B_s}^{0}}^{(0)}=-2 D_{a}-\frac{8}{3} E_{a}, \quad V_{{\bar B \bar
B_s}^{1/2}}^{(0)}=-2 D_{a}-\frac{8}{3} E_{a}.  \label{eqV0}
\end{eqnarray}

\begin{figure}[!htbp]
\centering
\scalebox{1}{\includegraphics{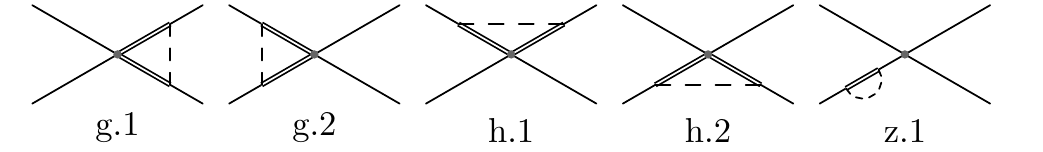}}\\
\caption{The loop diagrams with a contact vertex. The thin solid,
thick solid and dashed lines represent the heavy pseudoscalar
mesons, heavy vector mesons, and light pseudoscalar mesons
respectively.
 } \label{LoopCont}
 \end{figure}

\begin{figure}[!htbp]
\centering
\scalebox{1}{\includegraphics{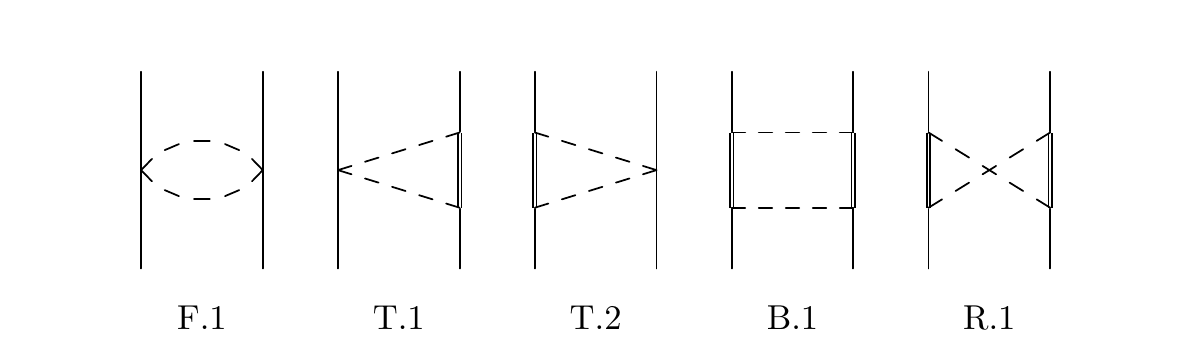}}\\
\caption{The 2$\phi$-exchange diagrams including the football
diagram (F.1), triangle diagrams (T.1 and T.2), box diagram (B.1),
and crossed diagram (R.1).
 } \label{LoopTPi}
 \end{figure}

The loop diagrams in Figs. \ref{LoopCont} and \ref{LoopTPi} will
contribute at the next to leading order. The diagrams h.1, h.2,
and B.1 in Figs. \ref{LoopCont} and \ref{LoopTPi} contain both 2PR
and 2PI parts if using the ordinary Feynman rules of HM$\chi$PT.
We must remove the 2PR contribution to get the correct potentials.
The 2PR parts result from the double poles of the two heavy
mesons, which can be removed by the careful subtraction in the
propagator of the heavy mesons
\begin{eqnarray}
&&   \frac{1}{[v\cdot p_1+\delta_1+i\varepsilon]
      [v\cdot p_2+\delta_2+i\varepsilon]}\nonumber
\\&= &
\left\{
           \begin{array}{ll}
 \displaystyle   \frac{1}{v\cdot p_1+\delta_1+i\varepsilon}[\frac{-1}{v\cdot p_1+\delta_1+i\varepsilon}-2\pi i \delta(v\cdot p_1+\delta_1)]
                \to \frac{-1}{(v\cdot p_1+\delta_1+i\varepsilon)^2}
                            &~  v\cdot p_2+\delta_2=-v\cdot p_1-\delta_1\\
 \displaystyle  \frac{1}{[v\cdot p_1+\delta_1+i\varepsilon]
      [v\cdot p_2+\delta_2+i\varepsilon]}
                            &~  {\rm other}
           \end{array}
    \right. .
\end{eqnarray}

We calculate these Feynman diagrams with dimension regularization
and modified minimal subtraction scheme. The divergent terms
proportional to $L$ will be absorbed by the contact terms at
$O(\epsilon^2)$, where
\begin{equation}
  L=\frac{\lambda^{D-4}}{16\pi^2}\left\{\frac1{D-4}+\frac12(\gamma_E-1-\ln 4\pi)\right\}.
\end{equation}
Here $\gamma_E$ is the Euler constant 0.5772157, $\lambda$ is the
scale of the dimension regularization, and we set $\lambda=4\pi
f$.

The potentials are finite after the renormalization of the
wavefunctions and vertices. The diagram z.1 in Fig. \ref{LoopCont}
arises from the renormalization of the wavefuntions. The combined
divergence generated by diagrams g.1 and g.2 in Fig.
\ref{LoopCont} can be absorbed by the LECs $D_{ai}^v$, $E_{ai}^v$,
$D_b^{h/d}$, and $E_b^{h/d}$. The divergence generated by the
diagram h.1 or h.2 can be absorbed by the redefinitions of LECs in
$\mathcal L^{(2,h)}_{4H}$, $\mathcal
L^{(2,d)}_{4H}$, and $\mathcal L^{(2,v)}_{4H}$. For the 2$\phi$
exchange diagrams in Fig. \ref{LoopTPi}, the divergence of the
football or triangle diagram can be absorbed by $E_{ai}^v$, and
$E_a^{h/d}$. The divergence of the box and crossed diagram can be
absorbed by $D_{ai}^v$, $D_a^{h/d}$, $E_{ai}^v$, and $E_a^{h/d}$.

The potentials obtained from the same Feynman diagram for the
different channels differ just by a flavor dependent coefficient.
So it is convenient to write down the potential of the channel
`$ch$' in the form
\begin{equation}
  V_{ch}^{(2)}=-\frac14\sum_{diag, m_1, m_2, ...}
         {\beta^{diag}}_{ch}(m_1, m_2,...)\otimes Y^{diag}(m_1, m_2, ...) ,  \label{Vch}
\end{equation}
where `$diag$' runs over all the Feynman diagrams shown in Figs.
\ref{LoopCont} and \ref{LoopTPi}, and $m_i$ runs over
$\{m_\pi,m_K,m_\eta\}$. $Y$ is a scalar function independent of
the flavor structure of the channel `$ch$', while $\beta$ is the
flavor dependent coefficient.

The corresponding $Y$ functions of Fig. \ref{LoopCont} are
\begin{eqnarray}
Y^{g.1}(m)&\equiv& -\frac{g^2}{f^2}\left\{[\frac{
D}{4}-\frac{1}{4}]J^g_{22}\right\}_r(m,\mathcal E-q_0-\Delta
,\mathcal E-\Delta ),\label{eqY1}
\\
Y^{g.2}(m)&\equiv& -\frac{g^2}{f^2}\left\{[\frac{
D}{4}-\frac{1}{4}]J^g_{22}\right\}_r(m,\mathcal E+q_0-\Delta
,\mathcal E-\Delta ),
\\
Y^{h.1}(m)&\equiv& -\frac{ g^2}{f^2}\left\{[\frac{
D}{4}-\frac{1}{4}]J^h_{22}\right\}_r(m,\mathcal E-\Delta ,\mathcal
E-\Delta ),
\\
Y^{h.2}(m)&\equiv& -\frac{ g^2}{f^2}\left\{[\frac{
D}{4}-\frac{1}{4}]J^h_{22}\right\}_r(m,\mathcal E-q_0-\Delta
,\mathcal E+q_0-\Delta ),
\\
Y^{z.1}(m)&\equiv& \left.\frac{g^2}{f^2}\left\{[\frac{
D}{4}-\frac{1}{4}]\partial_x
J^a_{22}\right\}_r(m,x)\right|_{x\to-\Delta},\label{eqY5}
\end{eqnarray}
where we work in the center-of-mass frame of the incoming heavy
mesons, $\mathcal E$ is the residual energy of the incoming heavy
meson (the difference between the energy and the $\bar B$ meson
mass), and $q$ is the transferred momentum. The definitions of the
$J$ functions are collected in Appendix \ref{secFunc}. $\{X\}_r$
represents the finite part of $X$,
\begin{equation}
   \{X\}_r=\lim_{D\to4} (X-L\frac{\partial}{\partial L} X)
                 +\frac1{16\pi^2}\lim_{D\to4}(\frac{\partial}{\partial D}\frac{\partial}{\partial L} X).
\end{equation}

The loop diagrams in Fig. \ref{LoopCont} are made up of the
contact vertices of $\mathcal L^{(0)}_{4H}$, so the flavor
dependent coefficient $\beta$ can be written down as
\begin{equation}
  \beta=D_{a}\tilde\beta_{\rm Da}+D_{b}\tilde\beta_{\rm Db}+E_{a}\tilde\beta_{\rm Ea}+E_{b}\tilde\beta_{\rm Eb}.
\end{equation}
Actually, only three coefficients $\beta$s for the diagrams g.1,
h.1, and z.1 in Fig.\ref{LoopCont} are independent. For
convenience, we list $\{\tilde\beta^{g.1}_{\rm
Da},\tilde\beta^{z.1}_{\rm Da},\tilde\beta^{g.1}_{\rm Db}
,\tilde\beta^{h.1}_{\rm Db}\}$ in Table~\ref{TabD} and
$\{\tilde\beta^{g.1}_{\rm Ea},\tilde\beta^{z.1}_{\rm
Ea},\tilde\beta^{g.1}_{\rm Eb} ,\tilde\beta^{h.1}_{\rm Eb}\}$ in
Table~\ref{TabE}. The coefficients $\beta$s for g.2 and h.2 can be
obtained by the following relations
\begin{equation}
  \tilde\beta^{g.2}_{\rm Da}=\tilde\beta^{g.1}_{\rm Da}, \quad
  \tilde\beta^{g.2}_{\rm Db}=\tilde\beta^{g.1}_{\rm Db}, \quad
  \tilde\beta^{h.2}_{\rm Db}=\tilde\beta^{h.1}_{\rm Db}, \quad
  \tilde\beta^{g.2}_{\rm Ea}=\tilde\beta^{g.1}_{\rm Ea}, \quad
  \tilde\beta^{g.2}_{\rm Eb}=\tilde\beta^{g.1}_{\rm Eb}, \quad
  \tilde\beta^{h.2}_{\rm Eb}=\tilde\beta^{h.1}_{\rm Eb},
\end{equation}
and all the others are zero.

\begin{table}[!htbp]
\caption{The coefficients $\beta$s of the loop diagrams with a
contact term: $\{\tilde\beta^{g.1}_{\rm Da},\tilde\beta^{z.1}_{\rm
Da},\tilde\beta^{g.1}_{\rm Db} ,\tilde\beta^{h.1}_{\rm Db}\}$
}\label{TabD}
\begin{tabular}{cccc}
\hline
0   &   ($m_\pi$)   &   ($m_K$) &   ($m_\eta$)  \\
\hline
${\bar B \bar B}^{1}$   &   $\displaystyle \{24,-48,-8,-8\}$    &   $\displaystyle \{16,-32,0,0\}$  &   $\displaystyle \left\{\frac{8}{3},-\frac{16}{3},-\frac{8}{3},-\frac{8}{3}\right\}$  \\
${\bar B \bar B}^{0}$   &   $\displaystyle \{24,0,-24,0\}$  &   $\displaystyle \{16,0,0,0\}$    &   $\displaystyle \left\{\frac{8}{3},0,\frac{8}{3},0\right\}$  \\
${\bar B_s \bar B_s}^{0}$   &   0   &   $\displaystyle \{32,-64,0,0\}$  &   $\displaystyle \left\{\frac{32}{3},-\frac{64}{3},-\frac{32}{3},-\frac{32}{3}\right\}$   \\
${\bar B \bar B_s}^{1/2}$   &   $\displaystyle \{12,-24,0,0\}$  &   $\displaystyle \{24,-48,-16,-16\}$  &   $\displaystyle \left\{\frac{20}{3},-\frac{40}{3},\frac{16}{3},\frac{16}{3}\right\}$ \\
\hline
\end{tabular}
\end{table}

\begin{table}[!htbp]
\caption{The coefficients $\beta$s of the loop diagrams with a
contact term: $\{\tilde\beta^{g.1}_{\rm Ea},\tilde\beta^{z.1}_{\rm
Ea},\tilde\beta^{g.1}_{\rm Eb} ,\tilde\beta^{h.1}_{\rm Eb}\}$
}\label{TabE}
\begin{tabular}{cccc}
\hline
0   &   ($m_\pi$)   &   ($m_K$) &   ($m_\eta$)  \\
\hline
${\bar B \bar B}^{1}$   &   $\displaystyle \left\{0,-64,-\frac{128}{3},-\frac{32}{3}\right\}$   &   $\displaystyle \left\{-\frac{32}{3},-\frac{128}{3},-32,0\right\}$   &   $\displaystyle \left\{\frac{32}{9},-\frac{64}{9},-\frac{32}{9},-\frac{32}{9}\right\}$    \\
${\bar B \bar B}^{0}$   &   $\displaystyle \{32,0,-32,0\}$  &   $\displaystyle \left\{-\frac{32}{3},0,-32,0\right\}$    &   $\displaystyle \left\{-\frac{64}{9},0,-\frac{64}{9},0\right\}$  \\
${\bar B_s \bar B_s}^{0}$   &   0   &   $\displaystyle \left\{-\frac{64}{3},-\frac{256}{3},-64,0\right\}$   &   $\displaystyle \left\{\frac{128}{9},-\frac{256}{9},-\frac{128}{9},-\frac{128}{9}\right\}$   \\
${\bar B \bar B_s}^{1/2}$   &   $\displaystyle \{-8,-32,-24,0\}$    &   $\displaystyle \left\{16,-64,-\frac{112}{3},-\frac{64}{3}\right\}$  &    $\displaystyle \left\{-\frac{136}{9},-\frac{160}{9},-\frac{152}{9},\frac{64}{9}\right\}$   \\
\hline
\end{tabular}
\end{table}

Now one can write down the potentials induced by Fig.
\ref{LoopCont} for the different channels. Let us take the
$\bar{B}\bar{B}^0$ system as an example. One can get the
contribution from the diagram g.1 by Tables \ref{TabD} and
\ref{TabE}
\begin{eqnarray}
  V_{\bar{B}\bar{B}^0}^{g.1}&=&-\frac{1}{4}\left[24D_a-24D_b+32E_a-32E_b\right]Y^{g.1}(m_{\pi})
    -\frac{1}{4}\left[16D_a-\frac{32}{3}E_a-32E_b\right]Y^{g.1}(m_K)
    \nonumber\\ &&
    -\frac{1}{4}\left[\frac{8}{3}D_a+\frac{8}{3}D_b-\frac{64}{9}E_a-\frac{64}{9}E_b\right]Y^{g.1}(m_{\eta}).
\end{eqnarray}

The diagrams in Fig. \ref{LoopTPi} represent the potentials with
the 2$\phi$ exchange. We list the $\beta$ functions
$\{\beta^{B.1},\beta^{R.1}\}$ in Table \ref{TabFTBR}, and all the
others can be obtained by the following relations
\begin{equation}
  \beta^{F.1}=\frac{-\beta^{B.1}+\beta^{R.1}}{16}, \quad \beta^{T.1}=\beta^{T.2}=\frac{-\beta^{B.1}+\beta^{R.1}}{4}. \label{eqbeta}
\end{equation}
The corresponding $Y$ functions of Fig. \ref{LoopTPi} are
\begin{eqnarray}
Y^{F.1}(m,M)&\equiv&
\frac{1}{f^4}\left\{[4]J^F_{22}+[q_0^2]J^F_{0}+[4
q_0^2]J^F_{11}+[4 q_0^2]J^F_{21}\right\}_r(m,M,q), \label{eqYF1}
\\
Y^{T.1}(m,M)&\equiv& \frac{g^2}{f^4}\left\{[D-1]J^T_{34}+[\frac{D
q_0}{2}-\frac{q_0}{2}]J^T_{21}+[D q_0-q_0]J^T_{31} +[-\vec
q^2]J^T_{24}+[-\vec q^2]J^T_{33} +[-\frac{1}{2} q_0 \vec
q^2]J^T_{11} \right.\nonumber\\ &&\left. \qquad +[-\frac{3}{2} q_0
\vec q^2]J^T_{22} +[-q_0 \vec q^2]J^T_{32}\right\}_r(m,M,\mathcal
E+q_0-\Delta ,q),
\\
Y^{T.2}(m,M)&\equiv& \frac{g^2}{f^4}\left\{[D-1]J^T_{34}+[\frac{D
q_0}{2}-\frac{q_0}{2}]J^T_{21}+[D q_0-q_0]J^T_{31} +[-\vec
q^2]J^T_{24}+[-\vec q^2]J^T_{33}+[-\frac{1}{2} q_0 \vec
q^2]J^T_{11} \right.\nonumber\\ &&\left. \qquad +[-\frac{3}{2} q_0
\vec q^2]J^T_{22} +[-q_0 \vec q^2]J^T_{32}\right\}_r(m,M,\mathcal
E-\Delta ,q),
\\
Y^{B.1}(m,M)&\equiv&
\frac{g^4}{f^4}\left\{[\frac{D^2}{4}-\frac{1}{4}]J^B_{41}+[-\frac{1}{4}
\vec q^2]J^B_{21} +[-\frac{1}{2} D \vec q^2-\frac{1}{2} \vec
q^2]J^B_{31}+[-\frac{1}{2} D \vec q^2-\frac{1}{2} \vec
q^2]J^B_{42} +[\frac{1}{4} (\vec q^2)^2]J^B_{22}
\right.\nonumber\\ &&\left. \qquad +[\frac{1}{2} (\vec
q^2)^2]J^B_{32} +[\frac{1}{4} (\vec
q^2)^2]J^B_{43}\right\}_r(m,M,\mathcal E-\Delta ,\mathcal E-\Delta
,q),
\\
Y^{R.1}(m,M)&\equiv&
\frac{g^4}{f^4}\left\{[\frac{D^2}{4}-\frac{1}{4}]J^R_{41}+[-\frac{1}{4}
\vec q^2]J^R_{21} +[-\frac{1}{2} D \vec q^2-\frac{1}{2} \vec
q^2]J^R_{31}+[-\frac{1}{2} D \vec q^2-\frac{1}{2} \vec
q^2]J^R_{42} +[\frac{1}{4} (\vec q^2)^2]J^R_{22}
\right.\nonumber\\ &&\left. \qquad +[\frac{1}{2} (\vec
q^2)^2]J^R_{32} +[\frac{1}{4} (\vec
q^2)^2]J^R_{43}\right\}_r(m,M,\mathcal E-\Delta ,\mathcal
E+q_0-\Delta ,q).  \label{eqYR1}
\end{eqnarray}

\begin{table}[!htbp]
\caption{The coefficients $\beta$s of the 2-$\phi$ exchange
diagrams: $\{\beta^{B.1},\beta^{R.1}\}$ }\label{TabFTBR}
\begin{tabular}{cccccccccc}
\hline
                           &($m_\pi$, $m_\pi$)        &($m_K$, $m_K$)            &($m_\eta$, $m_\eta$)                                         &($m_\pi$, $m_K$)         &($m_\pi$, $m_\eta$)                                        &   ($m_K$, $m_\eta$)                                           &($m_K$, $m_\pi$)         &($m_\eta$, $m_\pi$)                                        &($m_\eta$, $m_K$)                                           \\
\hline
${\bar B \bar B}^{1}$      &$\displaystyle \{-1,-5\}$ &$\displaystyle \{0,-4\}$  &$\displaystyle \left\{-\frac{1}{9},-\frac{1}{9}\right\}$     &0                        &$\displaystyle \left\{-\frac{1}{3},-\frac{1}{3}\right\}$   &   0                                                           &0                        &$\displaystyle \left\{-\frac{1}{3},-\frac{1}{3}\right\}$   &0                                                           \\
${\bar B \bar B}^{0}$      &$\displaystyle \{-9,3\}$  &$\displaystyle \{0,4\}$   &$\displaystyle \left\{-\frac{1}{9},-\frac{1}{9}\right\}$     &0                        &$\displaystyle \{1,1\}$                                    &   0                                                           &0                        &$\displaystyle \{1,1\}$                                    &0                                                           \\
${\bar B_s \bar B_s}^{0}$  &0                         &$\displaystyle \{0,-8\}$  &$\displaystyle \left\{-\frac{16}{9},-\frac{16}{9}\right\}$   &0                        &0                                                          &   0                                                           &0                        &0                                                          &0                                                           \\
${\bar B \bar B_s}^{1/2}$  &0                         &$\displaystyle \{-4,0\}$  &$\displaystyle \left\{-\frac{4}{9},-\frac{4}{9}\right\}$     &$\displaystyle \{0,-3\}$ &0                                                          &   $\displaystyle \left\{\frac{4}{3},-\frac{5}{3}\right\}$     &$\displaystyle \{0,-3\}$ &0                                                          &$\displaystyle \left\{\frac{4}{3},-\frac{5}{3}\right\}$     \\
\hline
\end{tabular}
\end{table}

Again, taking the $\bar{B}\bar{B}^0$ channel as an example, the
potential from the diagram B.1 of Fig. \ref{LoopTPi} reads
\begin{equation}
 V_{\bar{B}\bar{B}^0}^{B.1}=-\frac{1}{4}\left[-9Y^{B.1}(m_{\pi},m_{\pi})-\frac{1}{9}Y^{B.1}(m_{\eta},m_{\eta})
  +Y^{B.1}(m_{\pi},m_{\eta})+Y^{B.1}(m_{\eta},m_{\pi})\right].
\end{equation}

Finally, the potentials $V^{(2)}$ at $O(\epsilon^2)$ can be
obtained by summing the products of the corresponding $\beta$ and
$Y$ as in Eq. (\ref{Vch}).

\section{Numerical results and discussions}\label{secNum}

\subsection{Potentials of $\bar B\bar B$ Mesons}

We have calculated the potentials of $\bar B\bar B$ mesons up to
$O(\epsilon^2)$ for four independent channels. The $O(\epsilon^2)$
potentials $V^{(2)}$ contain two parts $V^{(2,{\rm cont})}$ and
$V^{(2,2\phi)}$, corresponding to Fig. \ref{LoopCont} and Fig.
\ref{LoopTPi} respectively. We will focus on the potentials with
$\mathcal E=q^0=0$ in the momentum space. After the Fourier
transformation, we can get the traditional potentials in the
coordinate space. The other parameters are listed as follows
\cite{PDG2012,Liu2011a,Ohki2008,Detmold2012}
\begin{equation}
 m_\pi=139~{\rm MeV}, \quad
 m_K=494~{\rm MeV},  \quad
 m_\eta=\sqrt{(4 m_K^2 - m_\pi^2)/3}, \quad
    \Delta=46~{\rm MeV},\quad
    f=92~{\rm MeV},\quad
   g=0.52,
\end{equation}
where the $\Delta$ is the mass difference between $\bar B$ and
$\bar B^*$ mesons, $f$ is the pion decay constant, and $g$ is the
axial coupling constant from the unquenched lattice QCD
simulation.

The potentials $V^{(0)}$ and $V^{(2,{\rm cont})}$ are both
generated by the contact vertices. They are independent of the
transferred momentum $|\vec q|$. They are $\delta (\vec{r})$-like
potentials in the coordinate space. From Eq. (\ref{eqV0}), we
notice that the terms proportional to $D_{b}$ and $E_{b}$ in
$\mathcal L^{(0)}_{4H}$ do not contribute to $V^{(0)}$. The
potential vanishes in the channel ${(\bar B \bar B)}^{0}$ at the
leading order. At the next to leading order, the situation is
different:
\begin{eqnarray}
&&V^{(2,{\rm cont})}_{{\bar B \bar B}^{1}}=
-0.32   E_a     -0.32   E_b ,\quad V^{(2,{\rm cont})}_{{\bar B
\bar B}^{0}}=   0.19    D_a     -0.02   D_b     -0.085  E_a
-0.36   E_b ,\nonumber\\&& V^{(2,{\rm cont})}_{{\bar B_s \bar
B_s}^{0}}=                           -0.53   E_a     -0.53   E_b
,\quad V^{(2,{\rm cont})}_{{\bar B \bar B_s}^{1/2}}=
-0.32   E_a     -0.32   E_b .   \label{eqV2cntBB}
\end{eqnarray}
All the terms in $\mathcal L^{(0)}_{4H}$ contribute to $V^{(2,{\rm
cont})}$. However, in the ${\bar B \bar B}^{1}$, ${\bar B_s \bar
B_s}^{0}$, and ${\bar B \bar B_s}^{1/2}$ channels, the
contributions proportional to $D_a$ or $D_b$ from different
diagrams in Fig. \ref{LoopCont} cancel each other. Roughly
speaking, the corrections $V^{(2,{\rm cont})}$ are small compared
with the leading order contribution. We also notice that
\begin{equation}
  \left|V^{(2,{\rm cont})}_{{\bar B_s \bar B_s}^{0}}\right|>\left|V^{(2,{\rm cont})}_{{\bar B \bar B_s}^{1/2}}\right|
     \approx \left|V^{(2,{\rm cont})}_{{\bar B \bar B}^{1}}\right|.
\end{equation}

As we have emphasized in the previous section, the finite parts of
the $O(\epsilon^2)$ LECs in Eqs. (\ref{eqL2h})-(\ref{eqL2q}) also contribute to the
potential while its divergent parts cancel the divergence from the
one-loop diagrams. Unfortunately we are unable to fix these LECs
because of the lack of experimental data. In the following analysis,
we focus on the behavior of the 2$\phi$-exchange potentials.

\begin{figure}[!htbp]
 \centering
 \subfigure{
 \scalebox{0.8}{\includegraphics{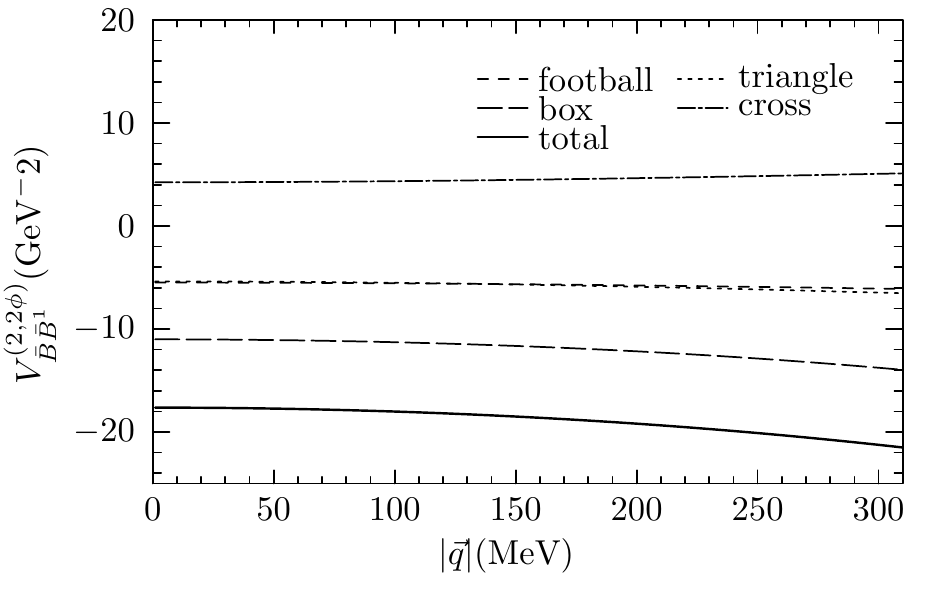} }
 }
 \subfigure{
 \scalebox{0.8}{\includegraphics{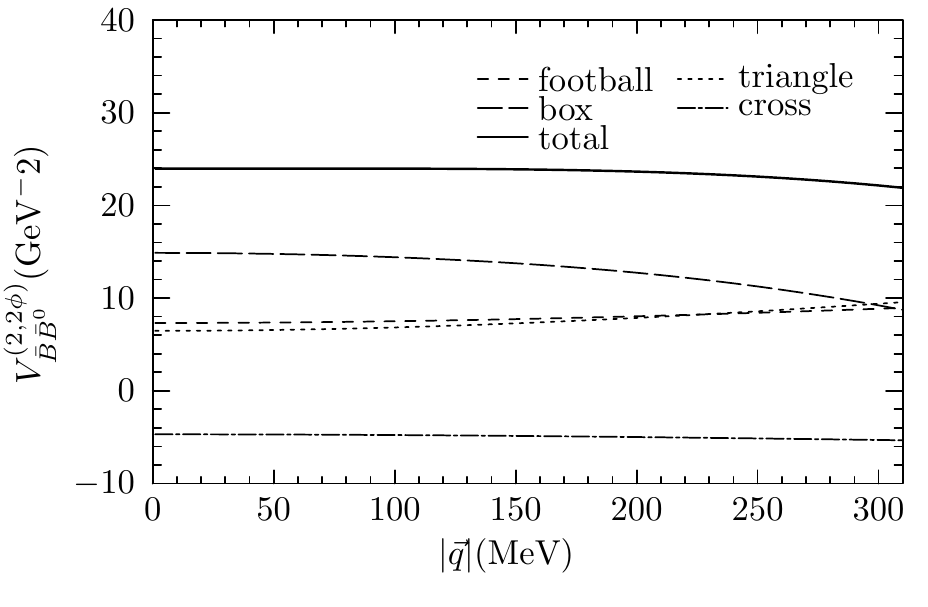} }
 }
  \subfigure{
 \scalebox{0.8}{\includegraphics{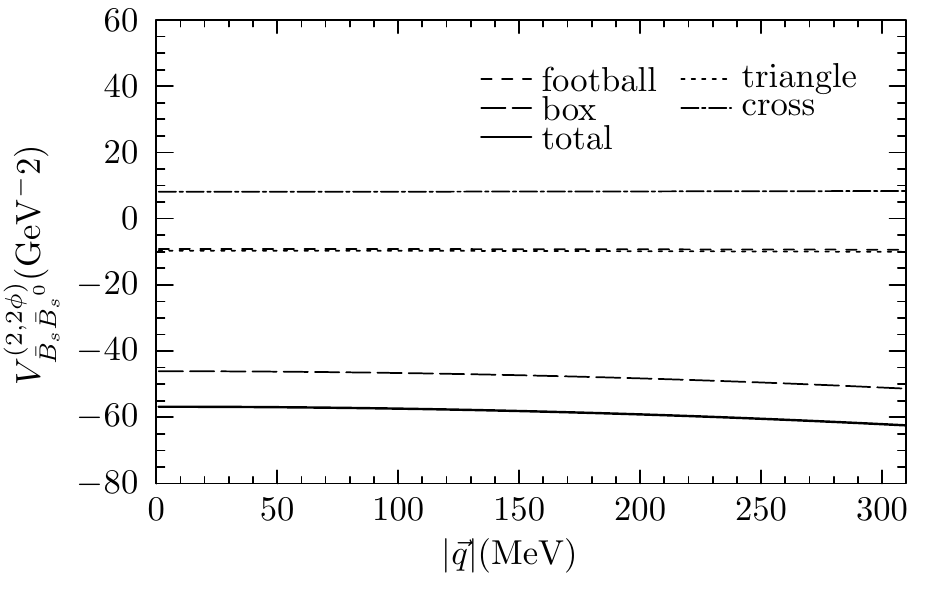} }
 }
 \subfigure{
 \scalebox{0.8}{\includegraphics{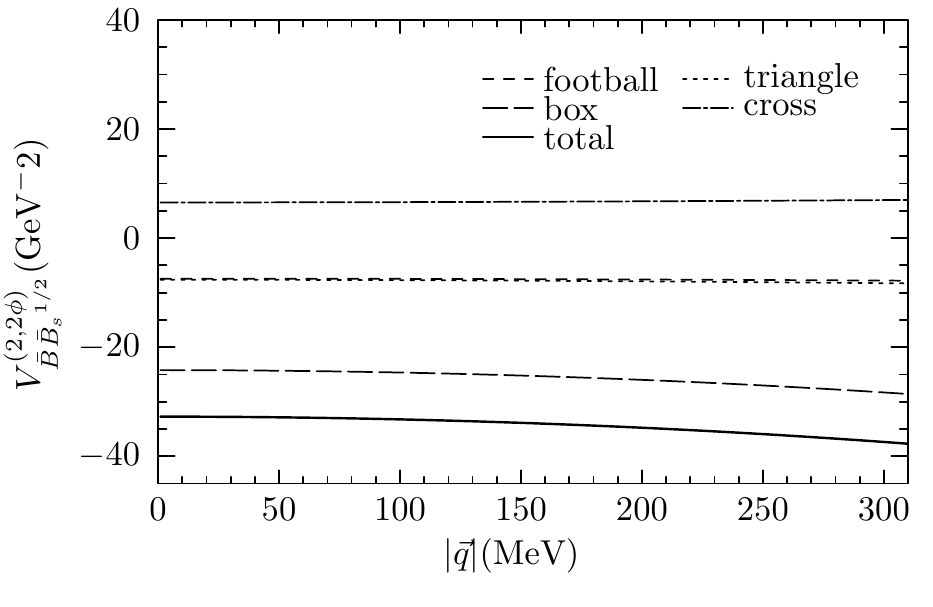} }
 }
 \caption{The $\bar B\bar B$ potentials with the 2$\phi$ exchange.}
 \label{figNumTPi}
 \end{figure}

We plot the 2$\phi$-exchange potentials $V^{(2,2\phi)}$ of the
$\bar B\bar B$ mesons in Fig. \ref{figNumTPi}. From the figure,
the contributions from the football and triangle diagrams are
coincidentally close in all the $\bar B\bar B$ channels. When the
transferred momentum is small, the sign of the potential from the
crossed diagrams is different from the other 2$\phi$-exchange
diagrams'. We have noticed that the
2$\phi$-exchange potentials of the ${\bar B \bar B}^{1}$, ${\bar
B_s \bar B_s}^{0}$, and ${\bar B \bar B_s}^{1/2}$ channels are
negative in the small-momentum region. In other words, the
2$\phi$-exchange interaction is attractive if we ignore contribution from
the LECs. In contrast, the 2$\phi$-exchange interaction of the
${\bar B \bar B}^{0}$ channel is repulsive in the small-momentum
region without these LECs. The 2$\phi$-exchange potential in the
${\bar B_s \bar B_s}^{0}$ channel is nearly twice larger than
those in the other channels.

The potentials from the football diagram, triangle diagram, box
diagram, and crossed diagram are proportional to $g^0$, $g^2$,
$g^4$, and $g^4$ respectively from Eqs.
(\ref{eqYF1})-(\ref{eqYR1}). So due to the coupling constant
$g=0.52$ is quite small, one would naively expect that the
potential from the triangle diagram is suppressed by a factor
about $0.27$, and the potential from the box or cross diagram is
suppressed by about $0.07$ compared with that from the football
diagram. However, we do not see the suppression in the Fig.
\ref{figNumTPi}. There is also an enhancement of the flavor
coefficient $\beta$ for the triangle, box, and crossed diagram
from Eq. (\ref{eqbeta}), which roughly compensates the suppression
of the small $g$. That is why we see neither the suppression due
to the small $g$ nor the enhancement of $\beta$ in the numerical
results. If we let $g\to 1$, the potential from the box or crossed
diagram would be much larger than the potential from the triangle
diagram, which would be larger than the potential from the
football diagram.

From Fig. \ref{figNumTPi}, we notice that the contribution from
the box diagram dominates the potential $V^{(2,2\phi)}$. From
Eq. (\ref{eqYR1}), we have
\begin{equation}
  15 \lessapprox\frac{Y^{B.1}(m_{\phi_1},m_{\phi_2})}{Y^{B.1}(m_\pi,m_\pi)}\lessapprox50, \label{eq2piVs2phi}
\end{equation}
where the intermediate meson pair $\phi_1\phi_2$ can be $\pi K$,
$\pi \eta$, $K K$, $K\eta$, or $\eta\eta$. So the potential from
the box diagram is dominated by the intermediate states with at
least one kaon or eta meson.

\subsection{Potentials of the $D D$ Mesons}

\begin{figure}[!htbp]
 \centering
 \subfigure{
 \scalebox{0.8}{\includegraphics{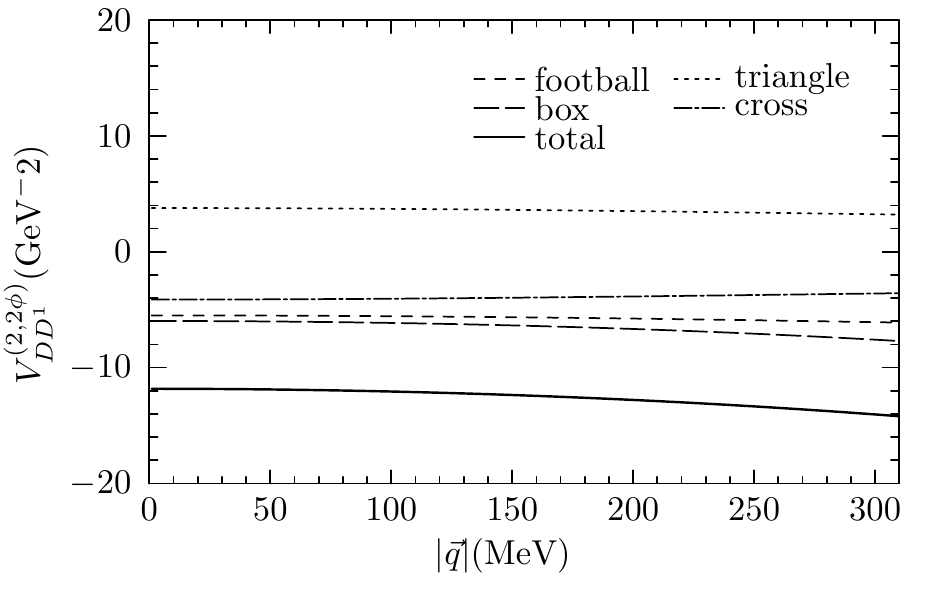} }
 }
 \subfigure{
 \scalebox{0.8}{\includegraphics{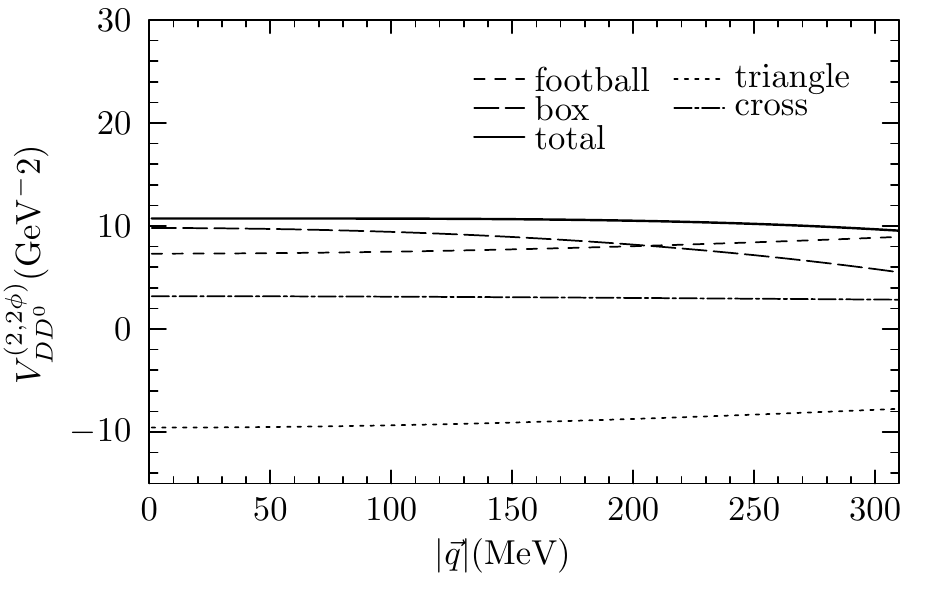} }
 }
  \subfigure{
 \scalebox{0.8}{\includegraphics{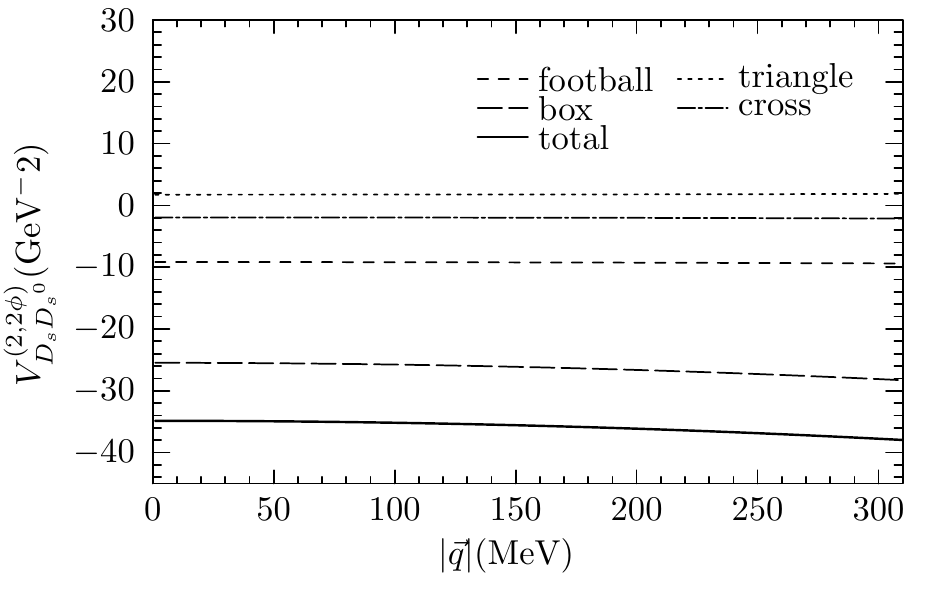} }
 }
 \subfigure{
 \scalebox{0.8}{\includegraphics{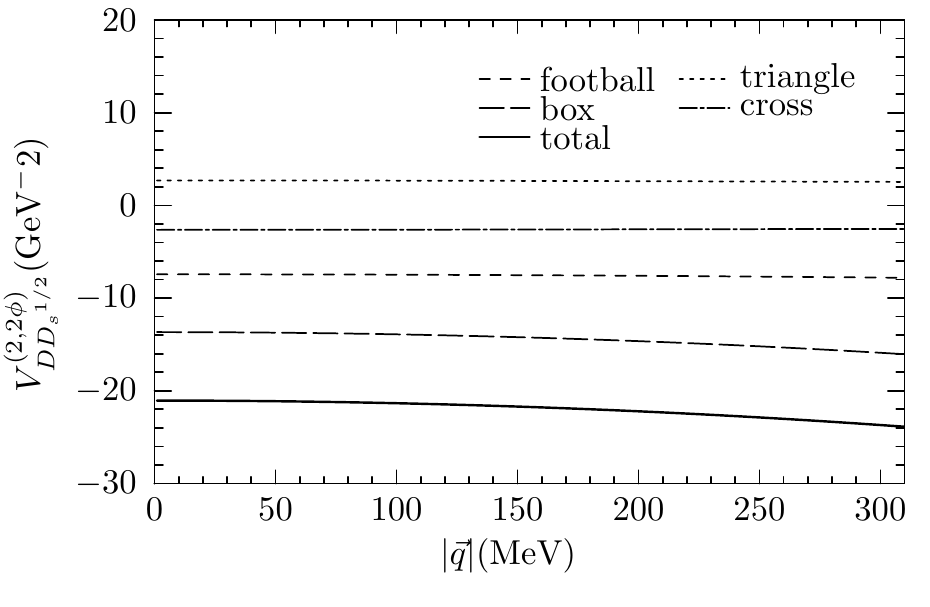} }
 }
 \caption{The $DD$ potentials with the 2$\phi$ exchange.}
 \label{figNumTPiDD}
 \end{figure}

Similarly we can study the potentials between the $D^0$, $D^+$,
and $D_s^+$ mesons. Now the intermediate heavy vector mesons are
$D^{*0}$, $D^{*+}$, and $D_s^{*+}$. The mass difference $\Delta$
increases to $142~{\rm MeV}$. The axial coupling constant $g=0.59$
from the decay width of $D^{*+}$ \cite{PDG2012}. The LECs $D_{a}$,
$E_{a}$ etc should be modified correspondingly. The expressions
for the $DD$ mesons are the same as those for the $\bar B\bar B$'s
except that the channels are ${DD}^{1}$, ${DD}^{0}$, ${D_s
D_s}^{0}$, and ${D D_s}^{1/2}$. The $DD$ potentials with the
2$\phi$ exchange are plotted in Fig. \ref{figNumTPiDD}. The
potentials related to the contact terms are
\begin{eqnarray}
&&V^{(2,{\rm cont})}_{{DD}^{1}}=                            0.89
E^{DD}_a    +   0.89    E^{DD}_b    ,\quad V^{(2,{\rm
cont})}_{{DD}^{0}}=  -0.59   D^{DD}_a    +   0.2 D^{DD}_b
-0.0056 E^{DD}_a    +   1.1 E^{DD}_b    ,\nonumber\\&& V^{(2,{\rm
cont})}_{{D_sD_s}^{0}}=                          1.1 E^{DD}_a    +
1.1 E^{DD}_b    ,\quad V^{(2,{\rm cont})}_{{DD_s}^{1/2}}=
1.1 E^{DD}_a    +   1.1 E^{DD}_b    .  \label{eqV2cntDD}
\end{eqnarray}

We notice that there is big difference between the potentials
$V^{(2,{\rm cont})}$ of the $\bar B\bar B$ and $DD$ mesons by
comparing Eqs. (\ref{eqV2cntBB}) and (\ref{eqV2cntDD}). The
difference originates from the different axial coupling $g$ and
mass difference $\Delta$. The signs of most of the terms in
$V^{(2,{\rm cont})}$ are different for the bottom and charm cases
if assuming $D^{DD}_{a/b}$ ($E^{DD}_{a/b}$) is equal to
$D_{a/b}$($E_{a/b}$) for the $\bar B$ mesons. One obtains the
relation
\begin{eqnarray}
\left|V^{(2,{\rm cont})}_{{DD_s}^{1/2}}\right|\approx
\left|V^{(2,{\rm cont})}_{{D_sD_s}^{0}}\right|>\left|V^{(2,{\rm
cont})}_{{DD}^{1}}\right|,
\end{eqnarray}
which is different from that in the case of the $\bar B$ mesons.

Comparing Fig. \ref{figNumTPi} and Fig. \ref{figNumTPiDD}, one can
find that the total $V^{(2,2\phi)}_{\bar B\bar B}$ is roughly
twice of $V^{(2,2\phi)}_{DD}$ for each channel. Moreover, the
separate contributions from the crossed or triangle diagrams have
opposite signs for the $\bar B$ and $D$ mesons.

One can obtain the potentials of anti-heavy mesons based on
C-parity conservation
\begin{equation}
  V_{{BB}^I}=V_{{\bar B\bar B}^I}, \quad
  V_{{\bar D \bar D}^I}=V_{{D D}^I}.
\end{equation}

\section{Comparison between Results in Different Schemes}\label{secCom}

One can also systematically study the potentials of the heavy
pseudoscalar mesons without heavy vector mesons as the explicit
degrees. The contributions from the heavy vector mesons, as well
as other resonances, will be embodied in the low-energy constants.
In the scheme without heavy vector mesons, the potential at the
leading order remains the same. However, only the football diagram
survives at $O(\epsilon^2)$.

It is also very interesting to investigate the case with strict
heavy quark spin symmetry. Now the heavy vector mesons are
included as the explicit degrees but the mass difference $\Delta$
is set to be zero. When $\Delta$ approaches 0, the potentials
induced by the diagrams h.1, h.2 and B.1 will approach infinity if
the 2PR contributions are not removed. The
two-heavy-vector-meson-reducible contribution does not appear when
the mass difference $\Delta$ is finite. It only appears as
$\Delta=0$ before the subtraction. So there is a jump in the
potentials as $\Delta$ goes from nonzero to zero. When solving the
nonperturbative equation to get the observable, the potentials
such as $V_{\bar B\bar B\to \bar B^*\bar B^*}$ must be included as
$\Delta=0$. But they are not necessary as $\Delta\neq 0$. Thus the
jump might vanish for the observable such as the binding energy.

In the new approximations, the potentials do not change at the
leading order. The difference appears at the next to leading
order. In the approximation without the heavy vector mesons, there
does not exist those diagrams in Fig. \ref{LoopCont}, so
$V^{(2,{\rm cont})}=0$. In the approximation including the heavy
vector mesons and $\Delta=0$, we have
\begin{eqnarray}
&&V^{(2,{\rm cont})}_{{\bar B \bar B}^{1}}=
-0.78   E_a     -0.78   E_b ,\quad V^{(2,{\rm cont})}_{{\bar B
\bar B}^{0}}=   0.48    D_a     -0.077  D_b     -0.17   E_a
-0.91   E_b ,\nonumber\\&& V^{(2,{\rm cont})}_{{\bar B_s \bar
B_s}^{0}}=                           -1.2    E_a     -1.2    E_b
,\quad V^{(2,{\rm cont})}_{{\bar B \bar B_s}^{1/2}}=
-0.88   E_a     -0.88   E_b , \label{eqV2cntBDlt}
\end{eqnarray}
and
\begin{eqnarray}
&&V^{(2,{\rm cont})}_{{DD}^{1}}=                            -1.0
E^{DD}_a        -1.0    E^{DD}_b    ,\quad V^{(2,{\rm
cont})}_{{DD}^{0}}=  0.62    D^{DD}_a        -0.099  D^{DD}_b
-0.22   E^{DD}_a        -1.2    E^{DD}_b    ,\nonumber\\&&
V^{(2,{\rm cont})}_{{D_sD_s}^{0}}=                          -1.6
E^{DD}_a        -1.6    E^{DD}_b    ,\quad V^{(2,{\rm
cont})}_{{DD_s}^{1/2}}=                          -1.1    E^{DD}_a
-1.1    E^{DD}_b    . \label{eqV2cntDDlt}
\end{eqnarray}
The difference between the $\bar B\bar B$ and $DD$ potentials only
originates from the different axial coupling $g$ as $\Delta=0$,
\begin{equation}
 \frac{ V^{(2,{\rm cont})}_{{\bar B \bar B}}}{V^{(2,{\rm cont})}_{{DD}}}=\frac{0.52^2}{0.59^2}=0.8, \quad \text{for~} \Delta=0.
\end{equation}

The $\bar B \bar B$ potential with $\Delta=0$ is nearly twice as
large as that with $\Delta=46$ MeV in every channel by comparing
Eqs. (\ref{eqV2cntBB}) and (\ref{eqV2cntBDlt}). The potentials
with $\Delta=46$ MeV are approximately equal in the channels ${\bar B
\bar B}^{1}$ and ${\bar B \bar B_s}^{1/2}$, while those with
$\Delta=0$ are not equal. The difference between the $DD$
potentials is even larger with different $\Delta$ from Eqs.
(\ref{eqV2cntDD}) and (\ref{eqV2cntDDlt}). The sign of the $DD$
potential as $\Delta=0$ is different from that as $\Delta=142$ MeV
in every channel. The potential of the channel ${D_sD_s}^{0}$ is
very close to that of the channel ${DD_s}^{1/2}$ for the case with
$\Delta=142$ MeV, but the situation is different for the case with
$\Delta=0$.

The difference between the potentials with different $\Delta$
mainly results from the subtraction of the
2-heavy-vector-meson-reducible contributions to get the potentials
of the heavy pseudoscalar mesons as $\Delta=0$. To recover the
2-heavy-vector-meson-reducible contributions, one should include
the potentials such as $V_{\bar B\bar B\to \bar B^*\bar B^*}$ when
solving the nonperturbative equations to get the observable as
$\Delta=0$.

\begin{figure}[!htbp]
 \centering
 \subfigure{
 \scalebox{0.8}{\includegraphics{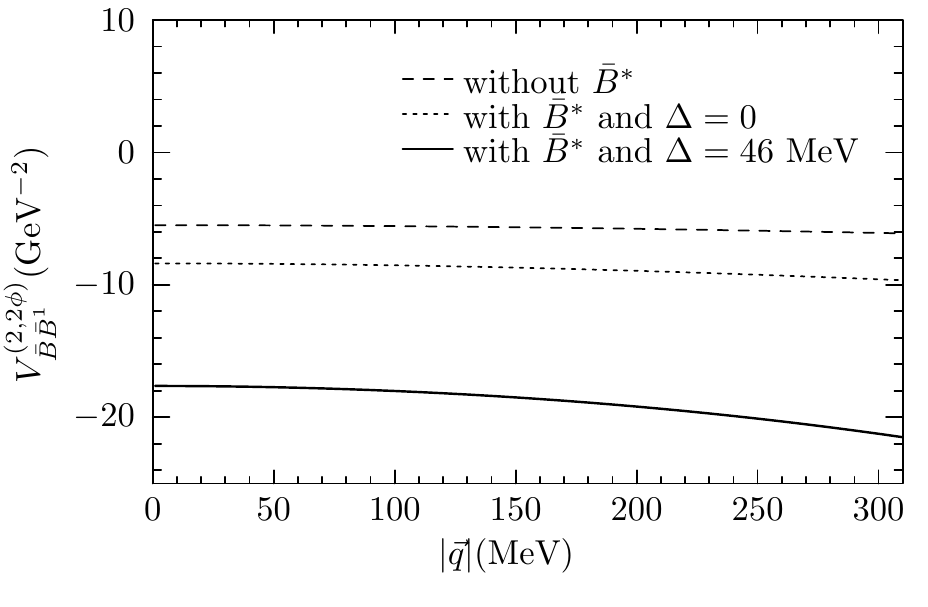} }
 }
 \subfigure{
 \scalebox{0.8}{\includegraphics{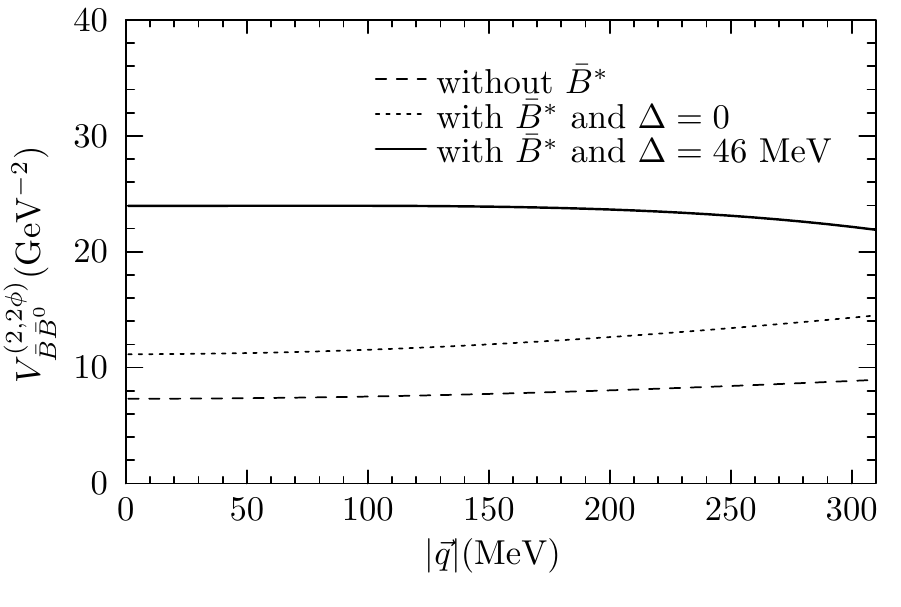} }
 }
  \subfigure{
 \scalebox{0.8}{\includegraphics{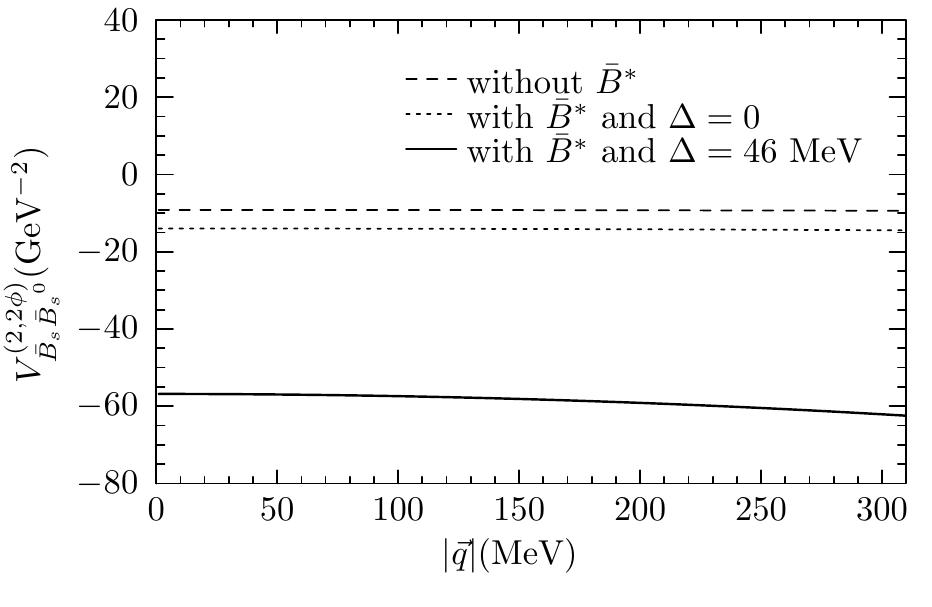} }
 }
 \subfigure{
 \scalebox{0.8}{\includegraphics{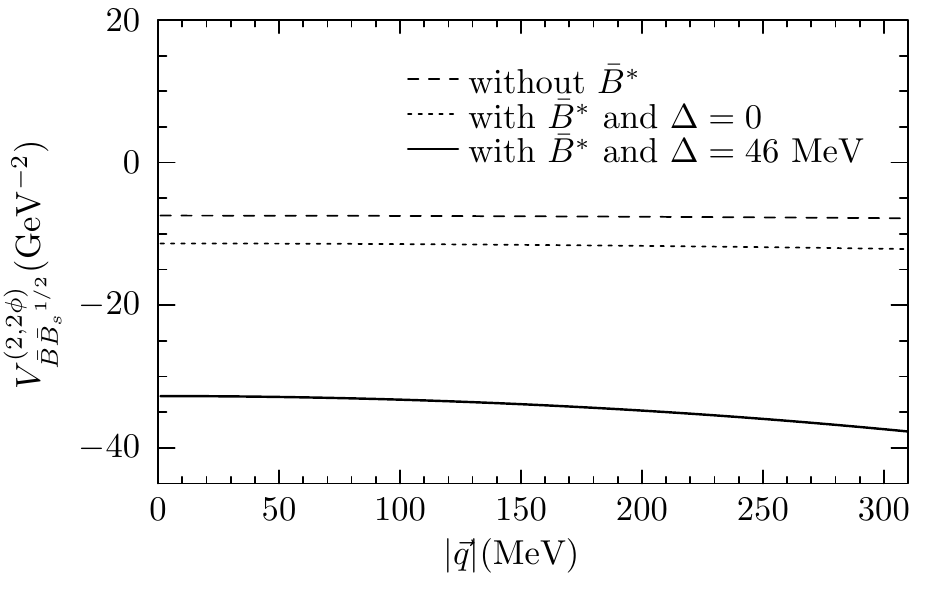} }
 }
 \caption{The $\bar B\bar B$ potentials with the 2$\phi$ exchange in different schemes.}
 \label{figBBappr}
 \end{figure}

\begin{figure}[!htbp]
 \centering
 \subfigure{
 \scalebox{0.8}{\includegraphics{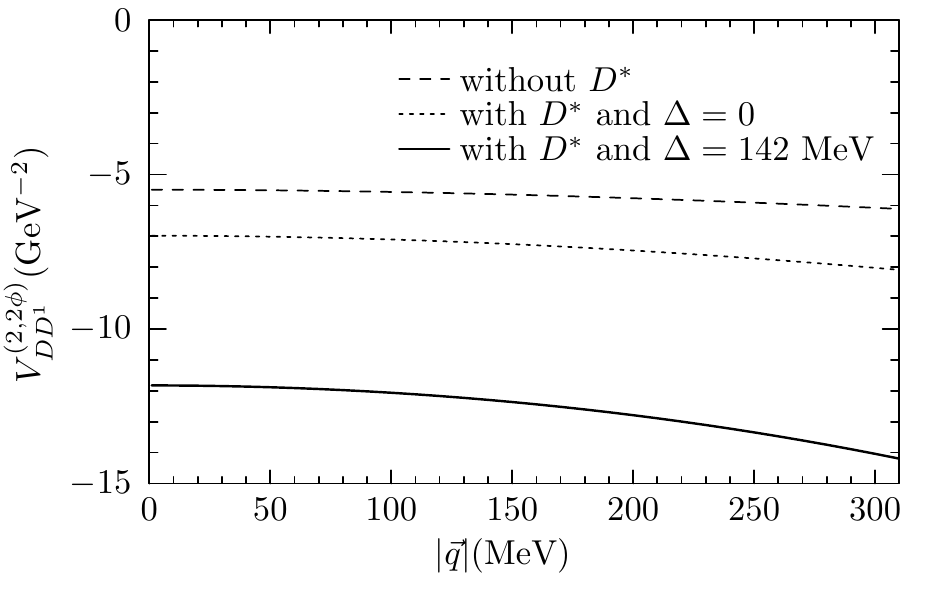} }
 }
 \subfigure{
 \scalebox{0.8}{\includegraphics{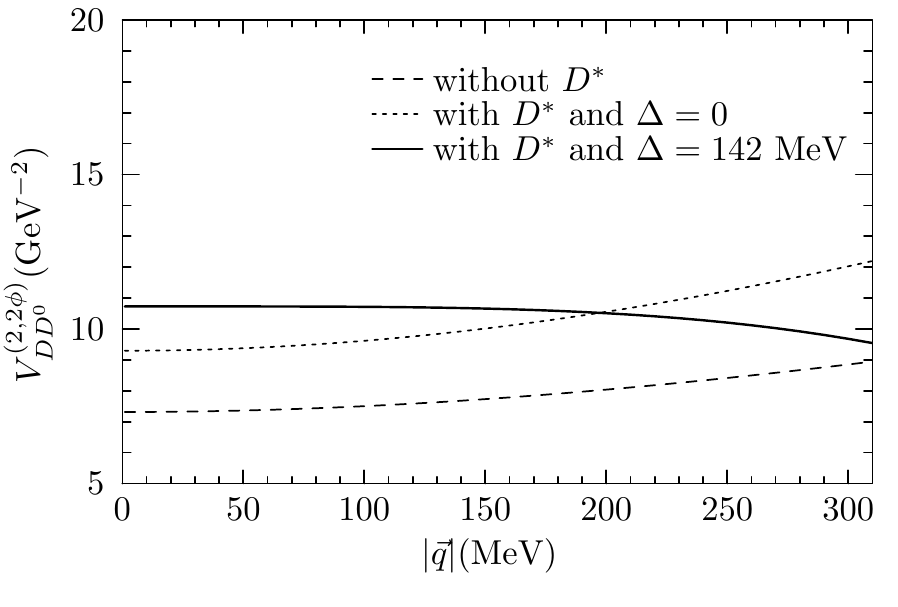} }
 }
  \subfigure{
 \scalebox{0.8}{\includegraphics{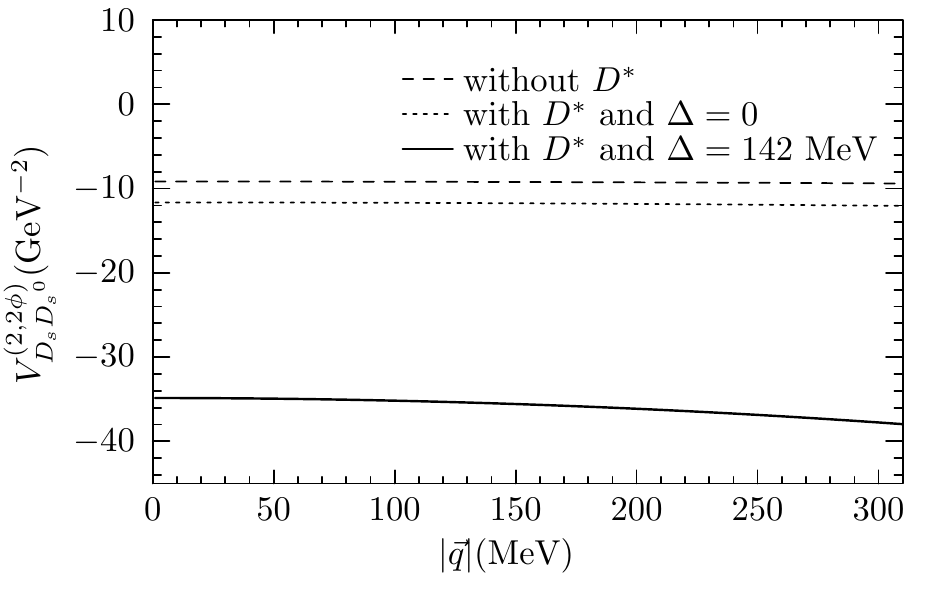} }
 }
 \subfigure{
 \scalebox{0.8}{\includegraphics{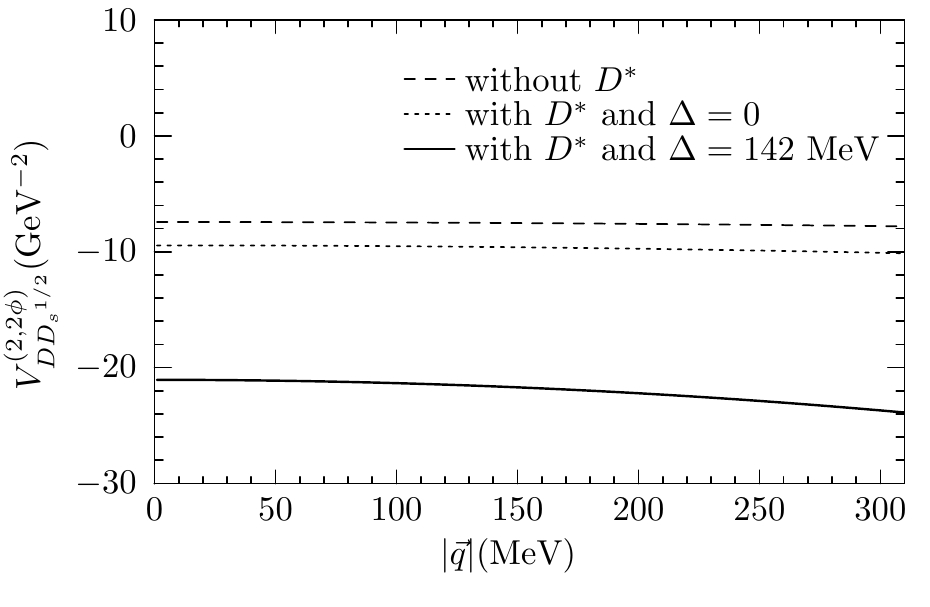} }
 }
 \caption{The $DD$ potentials with the 2$\phi$ exchange in different schemes.}
 \label{figDDappr}
 \end{figure}

We list the $\bar B\bar B$ and $DD$ potentials with 2$\phi$
exchange in different approximations in Figs. \ref{figBBappr} and
\ref{figDDappr} respectively. From Fig. \ref{figBBappr}, the
potentials are relatively close between the case without $\bar
B^*$ (case I) and that with $\Delta=0$ (case II). The potential
$|V|$ for case I or II is about 15\%-50\% of that for
the case with $\bar B^*$ and $\Delta=46~{\rm MeV}$ (case III) in
every channel. The dependence of the potential on $|\vec q|$ for
case III is slightly stronger than that for the other two cases in
the channels ${\bar B \bar B}^{1}$, ${\bar B_s \bar B_s}^{0}$, and
${\bar B \bar B_s}^{1/2}$. The potential decreases for case III
and increases for the other two cases as $|\vec q|$ grows in the
channel ${\bar B \bar B}^{0}$.

The situation in Fig. \ref{figDDappr} is similar to that in Fig.
\ref{figBBappr}. But the difference of the $DD$ potentials between
different cases is smaller than that of the $\bar B\bar B$
potentials. We notice that the $DD$ potential is equal to the
$\bar B\bar B$ potential for case I in each channel since the
potential from the football diagram is independent of the mass
difference $\Delta$ and the axial coupling constant $g$.

The 2$\phi$-exchange potentials in our results depend on the
energy scale $\lambda$ which arises from the dimensional
regularization. Thus the variations of the potentials with $\lambda$
might reveal the effect of the LECs to some extent. We reset
$\lambda=0.8 ~{\rm GeV}$ which is different from the previous one
$\lambda=4\pi f\approx 1.2 ~{\rm GeV}$. For comparison, we list
$V^{(2,{\rm 2\phi})}_{\bar B \bar B}$ potentials (for case III) with
different $\lambda$ in Table \ref{tabLam}. From Table \ref{tabLam},
one notices that as $\lambda$ goes from $4\pi f$ to 0.8 GeV, the
2$\phi$-exchange potential changes about 10\%$\sim$20\%.

\begin{table}[!htbp]
\caption{2$\phi$-exchange potentials of $\bar B \bar B$ with different $\lambda$ in units of ${\rm GeV^{-2}}$.   }\label{tabLam}
\begin{tabular}{c|cc|cc|cc|cc}
\hline
&   \multicolumn{2}{c|}{$V^{(2,{\rm 2\phi})}_{{\bar B \bar B}^{1}}$}
&     \multicolumn{2}{c|}{$V^{(2,{\rm 2\phi})}_{{\bar B \bar B}^{0}}$}
&\multicolumn{2}{c|}{$V^{(2,{\rm 2\phi})}_{{\bar B_s \bar B_s}^{0}}$}
 &\multicolumn{2}{c}{$V^{(2,{\rm 2\phi})}_{{\bar B \bar B_s}^{1/2}}$}\\
$|\vec q| (${\rm MeV}$)$ &0 & 300   &0 & 300 &0 & 300  &0 & 300 \\
\hline
$\lambda=4\pi f$   &-18. & -21.   &24. &22. &-57. &-62. &-33. &-37.\\
$\lambda=0.8~{\rm GeV}$&-14. &-18. &20. &18. &-51. &-56. &-28.&-32.\\
\hline
\end{tabular}
\end{table}

\section{Summary}\label{secSum}

In a short summary, we have calculated the potentials of the heavy
pseudoscalar mesons up to $O(\epsilon^2)$ in the momentum space
with HM$\chi$PT. We have carefully analyzed the tree-level
contribution and one-loop correction to the contact vertices, and
the 2$\pi$-exchange contribution. We have also discussed the
potentials in different schemes.

Generally speaking, the potential of hadrons can be separated into
the long-range, medium-range, and short-range parts. For the two
heavy pseudoscalar mesons, there does not exist the long-range
1$\phi$-exchange potential. The medium-range potential contains
the 2$\phi$-exchange potential and the contributions by the
Lagrangians (\ref{eqL2h})-(\ref{eqL2q}). The 2$\phi$-exchange
potential is model-independent (still renormalization-scheme dependent) since there are no unknown
constants in it, which is very essential for the medium-range
interaction of the heavy pseudoscalar mesons. The interaction
induced only by the 2$\phi$ exchange is repulsive in the channels
${\bar B \bar B}^{I=0}$, ${DD}^{I=0}$, while attractive in the
other channels. Unfortunately the leading order coupling constants
from the contact terms and LECs at $O(\epsilon^2)$ remain
undetermined due to lack of experimental data.

Once these LECs are extracted from lattice QCD simulation,
other model approaches or future experimental measurements, the
potentials derived in this work can be used to study the possible
molecular states or scattering phase shift of the two heavy
pseudoscalar mesons system. On the other hand, the analytical
chiral structures of the potentials of the heavy meson pair may be
useful in the extrapolation of the heavy meson interaction from
lattice QCD simulation.

\section{Acknowledgements}\label{secAck}

This project was supported by the National Natural Science
Foundation of China under Grants 11075004, 11021092, 11261130311
and Ministry of Science and Technology of China (2009CB825200).
This work is also supported in part by the DFG and the NSFC
through funds provided to the sino-germen CRC 110 ``Symmetries and
the Emergence of Structure in QCD''.

\appendix
\section{Some functions used for potentials}\label{secFunc}

The $J$ functions can be obtained by calculating the following
integrals in $D$ dimensions
\begin{eqnarray}
&&i\int\frac{d^D l \lambda^{4-D} }{ {(2\pi)}^D } \frac{ \{1,~
l^\alpha,~ l^\alpha l^\beta,~ l^\alpha l^\beta l^\gamma\} }
{(v\cdot l+\omega+i\varepsilon)(l^2-m^2+i\varepsilon)}
\nonumber\\&\equiv& \left\{J^a_0,\quad v^\alpha J^a_{11}, \quad
v^\alpha v^\beta J^a_{21}+g^{\alpha\beta}J^a_{22},\quad (g\vee
v)J^a_{31}+v^\alpha v^\beta v^\gamma J^a_{32} \right\} (m,\omega),
\\&&
i\int\frac{d^D l \lambda^{4-D} }{ {(2\pi)}^D } \frac{ \{1,~
l^\alpha,~ l^\alpha l^\beta\} }{(l^2-m^2+i\varepsilon)}
\nonumber\\&\equiv& \left\{J^c_0,\quad 0,\quad g^{\alpha
\beta}J^c_{21} \right\}(m),
\\&&
i\int\frac{d^D l \lambda^{4-D} }{ {(2\pi)}^D } \frac{ \{1,~
l^\alpha,~ l^\alpha l^\beta,~ l^\alpha l^\beta l^\gamma\} }
{(v\cdot l+\omega+i\varepsilon)[(+/-)v\cdot
l+\delta+i\varepsilon]_s(l^2-m^2+i\varepsilon) }
\nonumber\\&\equiv& \left\{J^{g/h}_0,\quad v^\alpha J^{g/h}_{11},
\quad v^\alpha v^\beta
J^{g/h}_{21}+g^{\alpha\beta}J^{g/h}_{22},\quad (g\vee
v)J^{g/h}_{31}+v^\alpha v^\beta v^\gamma J^{g/h}_{32} \right\}
(m,\omega,\delta),
\\&&
i\int\frac{d^D l \lambda^{4-D} }{ {(2\pi)}^D } \frac{ \{1,~
l^\alpha,~ l^\alpha l^\beta,~ l^\alpha l^\beta l^\gamma\} }
{(l^2-m^2+i\varepsilon)[(q+l)^2-M^2+i\varepsilon] }
\nonumber\\&\equiv& \left\{ J^F_0, \quad q^\alpha J^F_{11},\quad
q^\alpha q^\beta J^F_{21}+g^{\alpha\beta}J^F_{22},\quad (g\vee
q)J^F_{31}+q^\alpha q^\beta q^\gamma J^F_{32} \right\}(m,M,q),
\\&&
i\int\frac{d^D l \lambda^{4-D} }{ {(2\pi)}^D } \frac{ \{1,~
l^\alpha,~ l^\alpha l^\beta,~ l^\alpha l^\beta l^\gamma,~
l^\alpha l^\beta l^\gamma l^\delta\} } {(v\cdot
l+\omega+i\varepsilon)(l^2-m^2+i\varepsilon)[(q+l)^2-M^2+i\varepsilon]
} \nonumber\\&\equiv& \left\{ J^T_0, \quad q^\alpha
J^T_{11}+v^\alpha J^T_{12},\quad g^{\alpha \beta}
J^T_{21}+q^\alpha q^\beta J^T_{22}+v^\alpha v^\beta
J^T_{23}+(q\vee v)J^T_{24},\right. \nonumber\\&& (g\vee
q)J^T_{31}+q^\alpha q^\beta q^\gamma J^T_{32} +(q^2\vee
v)J^T_{33}+(g\vee v)J^T_{34}+(q\vee v^2)J^T_{35}+v^\alpha v^\beta
v^\gamma J^T_{36}, \nonumber\\&& (g\vee g)J^T_{41}+(g\vee
q^2)J^T_{42}+q^\alpha q^\beta q^\gamma q^\delta J^T_{43} +(g\vee
v^2)J^T_{44} + v^\alpha v^\beta v^\gamma v^\delta J^T_{45}
+(q^3\vee v) J^T_{46}+(q^2\vee v^2)J^T_{47} +(q\vee v^3) J^T_{48}
\nonumber\\&& \left. +(g\vee q\vee v)
J^T_{49}\right\}(m,M,\omega,q),
\\&&
i\int\frac{d^D l \lambda^{4-D} }{ {(2\pi)}^D } \frac{ \{1,~
l^\alpha,~ l^\alpha l^\beta,~ l^\alpha l^\beta l^\gamma,~
l^\alpha l^\beta l^\gamma l^\delta\} } {(v\cdot
l+\omega+i\varepsilon)[(+/-)v\cdot
l+\delta+i\varepsilon]_s(l^2-m^2+i\varepsilon)[(q+l)^2-M^2+i\varepsilon]
} \nonumber\\&\equiv& \left\{ J^{R/B}_0, \quad q^\alpha
J^{R/B}_{11}+v^\alpha J^{R/B}_{12},\quad g^{\alpha \beta}
J^{R/B}_{21}+q^\alpha q^\beta J^{R/B}_{22}+v^\alpha v^\beta
J^{R/B}_{23}+(q\vee v)J^{R/B}_{24},\right. \nonumber\\&& (g\vee
q)J^{R/B}_{31}+q^\alpha q^\beta q^\gamma J^{R/B}_{32} +(q^2\vee
v)J^{R/B}_{33}+(g\vee v)J^{R/B}_{34}+(q\vee
v^2)J^{R/B}_{35}+v^\alpha v^\beta v^\gamma J^{R/B}_{36},
\nonumber\\&& (g\vee g)J^{R/B}_{41}+(g\vee
q^2)J^{R/B}_{42}+q^\alpha q^\beta q^\gamma q^\delta J^{R/B}_{43}
+(g\vee v^2)J^{R/B}_{44} + v^\alpha v^\beta v^\gamma v^\delta
J^{R/B}_{45} +(q^3\vee v) J^{R/B}_{46}+(q^2\vee v^2)J^{R/B}_{47}
\nonumber\\&& \left.+(q\vee v^3) J^{R/B}_{48}+(g\vee q\vee v)
J^{R/B}_{49}
 \right\}(m,M,\omega,\delta,q),
\end{eqnarray}
where we have used the following Feynman rule for propagators of
two heavy mesons in HM$\chi$PT to remove the 2PR contributions
from the Feynman diagrams
\begin{eqnarray}
&&\frac{1}{(v\cdot l+\omega+i\varepsilon)[({\rm sgn})v\cdot
l+\delta+i\varepsilon]_s}\nonumber
\\&\equiv&
\left\{
           \begin{array}{ll}
 \displaystyle   \frac{1}{(v\cdot l+\omega+i\varepsilon)}[(-)\frac{1}{v\cdot l+\omega+i\varepsilon}-2\pi i\delta(v\cdot l+\omega)]_s
                \equiv (-)\frac{1}{(v\cdot l+\omega+i\varepsilon)^2}
                            &\quad  {\rm sgn}=-,~ \delta= -\omega\\
 \displaystyle  \frac{1}{(v\cdot l+\omega+i\varepsilon)[({\rm sgn})v\cdot l+\delta+i\varepsilon]}
                            &\quad  {\rm other}
           \end{array}
    \right. \label{eqPoleS}.
\end{eqnarray}
The notation $X \vee Y \vee Z \vee...$ represents the symmetrized
tensor of $X^\alpha Y^\beta Z^\gamma...$, and in detail,
\begin{eqnarray}
&&q \vee v \equiv q^\alpha v^\beta+q^\beta v^\alpha, \quad g \vee
q \equiv
g^{\alpha\beta}q^\gamma+g^{\alpha\gamma}q^\beta+g^{\gamma\beta}q^\alpha,
\quad g \vee v \equiv
g^{\alpha\beta}v^\gamma+g^{\alpha\gamma}v^\beta+g^{\gamma\beta}v^\alpha,
\quad \nonumber\\&& q^2 \vee v \equiv q^{\beta } q^{\gamma }
v^{\alpha }+q^{\alpha }
   q^{\gamma } v^{\beta }+q^{\alpha } q^{\beta } v^{\gamma }, \quad
q \vee v^2 \equiv q^{\gamma } v^{\alpha }
   v^{\beta }+q^{\beta } v^{\alpha } v^{\gamma }+q^{\alpha } v^{\beta } v^{\gamma },\quad
\nonumber\\&& g \vee g \equiv g^{\alpha \beta } g^{\gamma \delta
}+g^{\alpha \delta } g^{\beta \gamma }+g^{\alpha \gamma } g^{\beta
\delta }, \quad g \vee q^2 \equiv q^{\alpha } q^{\beta } g^{\gamma
\delta }+q^{\alpha } q^{\delta } g^{\beta \gamma } +q^{\alpha}
q^{\gamma } g^{\beta \delta }+q^{\gamma } q^{\delta } g^{\alpha
\beta } +q^{\beta } q^{\delta } g^{\alpha \gamma } +q^{\beta }
q^{\gamma } g^{\alpha \delta }, \quad \nonumber\\&& g \vee v^2
\equiv v^{\alpha } v^{\beta } g^{\gamma \delta } +v^{\alpha }
v^{\delta } g^{\beta \gamma }+v^{\alpha } v^{\gamma } g^{\beta
\delta }
   +v^{\gamma } v^{\delta } g^{\alpha \beta }+v^{\beta } v^{\delta   } g^{\alpha \gamma }
   +v^{\beta } v^{\gamma } g^{\alpha \delta }, \quad
\nonumber\\&& q^3\vee v \equiv q^{\beta } q^{\gamma } q^{\delta }
v^{\alpha }+q^{\alpha } q^{\gamma } q^{\delta} v^{\beta }
+q^{\alpha } q^{\beta } q^{\delta } v^{\gamma }+q^{\alpha }
q^{\beta } q^{\gamma } v^{\delta } ,\quad q\vee v^3 \equiv
q^{\delta } v^{\alpha } v^{\beta } v^{\gamma }+q^{\gamma }
v^{\alpha } v^{\beta } v^{\delta } +q^{\beta } v^{\alpha }
v^{\gamma } v^{\delta }+q^{\alpha } v^{\beta } v^{\gamma }
v^{\delta }, \nonumber\\&& q^2 \vee v^2 \equiv q^{\gamma }
q^{\delta } v^{\alpha } v^{\beta }+q^{\beta } q^{\delta }
v^{\alpha } v^{\gamma } +q^{\alpha } q^{\delta } v^{\beta }
v^{\gamma }+q^{\beta } q^{\gamma } v^{\alpha } v^{\delta }
+q^{\alpha } q^{\gamma } v^{\beta }  v^{\delta }+q^{\alpha }
q^{\beta } v^{\gamma } v^{\delta }, \nonumber\\&& g\vee q \vee v
\equiv q^{\beta } v^{\alpha } g^{\gamma \delta }+q^{\alpha }
v^{\beta } g^{\gamma \delta } +q^{\delta } v^{\alpha } g^{\beta
\gamma }+q^{\gamma } v^{\alpha } g^{\beta \delta }+q^{\alpha }
v^{\delta } g^{\beta \gamma } +q^{\alpha } v^{\gamma } g^{\beta
\delta}+q^{\delta } v^{\gamma } g^{\alpha \beta }+q^{\delta }
v^{\beta } g^{\alpha \gamma } +q^{\gamma } v^{\delta } g^{\alpha
\beta  } \nonumber\\&& \qquad\qquad\quad+q^{\gamma } v^{\beta }
g^{\alpha \delta }+q^{\beta } v^{\delta } g^{\alpha \gamma }
+q^{\beta } v^{\gamma } g^{\alpha \delta }.
\end{eqnarray}

\section{Estimation of contributions from the LECs at $O(\epsilon^2)$}\label{secFit}

We will estimate the contributions of the LECs  to the potentials at order $O(\epsilon^2)$, based on the data from the quenched lattice
QCD study in the two-flavor case \cite{Detmold2007}. In the quenched QCD, the ``quark-flow approach'' or ``quenched chiral perturbation theory'' should be used to get the potential\cite{Kilcup1990,Bernard1992,Labrenz1996}. In what follows, we apply for the quark-flow approach in which one uses the ordinary chiral Lagrangians but should  eliminate all diagrams containing virtual quark loops. As an estimation, we roughly consider the contributions of the tree diagrams to fit the data of the quenched lattice QCD study.

To $O(\epsilon^2)$, the potential from the tree diagrams in the three-flavor case can be written as
\begin{equation}
  V^{\rm tree}_{ch}(\vec q^2)=\frac{\alpha^{ch}}{\Lambda_0^2}+\frac{\beta_1^{ch}}{\Lambda_0^4} m_{\pi}^2+\frac{\beta_2^{ch}}{\Lambda_0^4} m_{K}^2+\frac{\beta_3^{ch}}{\Lambda_0^4} m_{\eta}^2+\frac{\gamma^{ch}}{\Lambda_0^4} \vec q^2,
  \quad \Lambda_0=1~{\rm GeV}, \label{eqVQQCD}
\end{equation}
where $\alpha^{ch}$ can be obtained by Eq. (14),
and $\beta_i^{ch}$ and $\gamma^{ch}$ are the linear combinations of LECs at $O(\epsilon^2)$.
In quenched QCD, the potential in the two-flavor case can be estimated as follows
\begin{equation}
  V^{\rm QQCD}_{{\bar B \bar B}^{I}}(\vec q^2)=\frac{a^I}{\Lambda_0^2}+\frac{b^I}{\Lambda_0^4} m_{\pi}^2+\frac{c^I}{\Lambda_0^4} \vec q^2,
  \quad \Lambda_0=1~{\rm GeV}, \label{eqVQQCD}
\end{equation}
where $a^I$, $b^I$, and $c^I$ are similar to $\alpha^{ch}$, $\beta_i^{ch}$, and $\gamma^{ch}$ in the three-flavor case.

From Eq. (\ref{eqV0}) one obtains $a^0=0$. Introducing Gaussian form factor  $\exp(-\vec q^2/\Lambda_G^2)$, we perform the Fourier Tansformation to get the potential in the coordinate space.

\begin{equation}
   V(\vec r)=\frac{1}{(2\pi)^3}\int d^3\vec q [\frac14V(\vec q^2)]e^{-\vec q^2/\Lambda_G^2}e^{-i \vec q\cdot \vec r}.\label{eqVq2r}
\end{equation}
The specific expression of the potential is
\begin{equation}
  V^{\rm QQCD}_{{\bar B \bar B}^{I}}(r)=
  \frac14 \frac{\Lambda_G^3 \exp(-\frac14 \Lambda_G^2 r^2)}{32\pi^{3/2}}
  [4 (\frac{a^I}{\Lambda_0^2}+\frac{b^I}{\Lambda_0^4} m_{\pi}^2) +\frac{c^I}{\Lambda_0^4}\Lambda_G^2(6-\Lambda_G^2 r^2)].
\end{equation}

\begin{figure}[!htbp]
 \centering
 \subfigure[$I=0$]{
 \scalebox{0.42}{\includegraphics{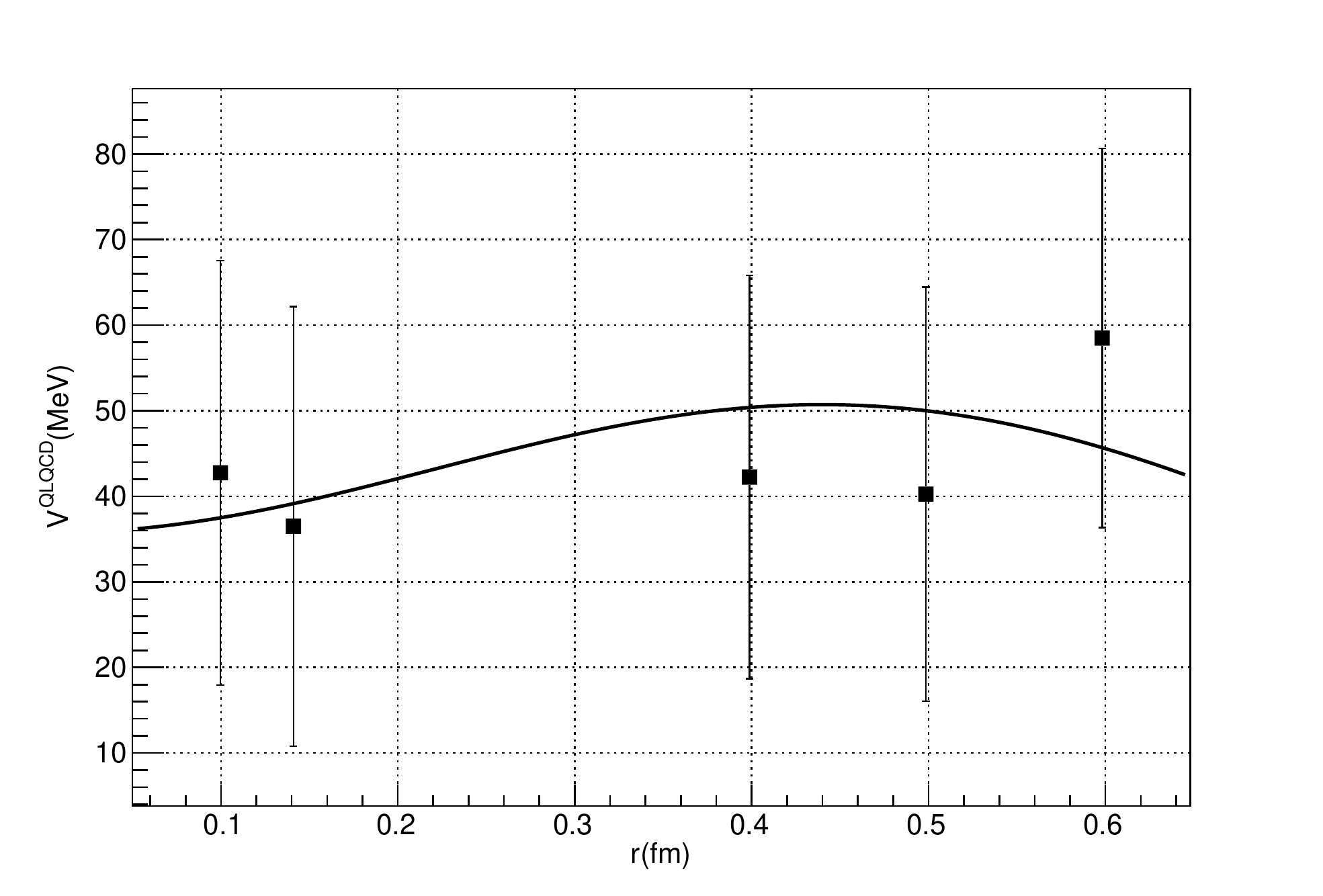} }
 }
 \subfigure[$I=1$]{
 \scalebox{0.42}{\includegraphics{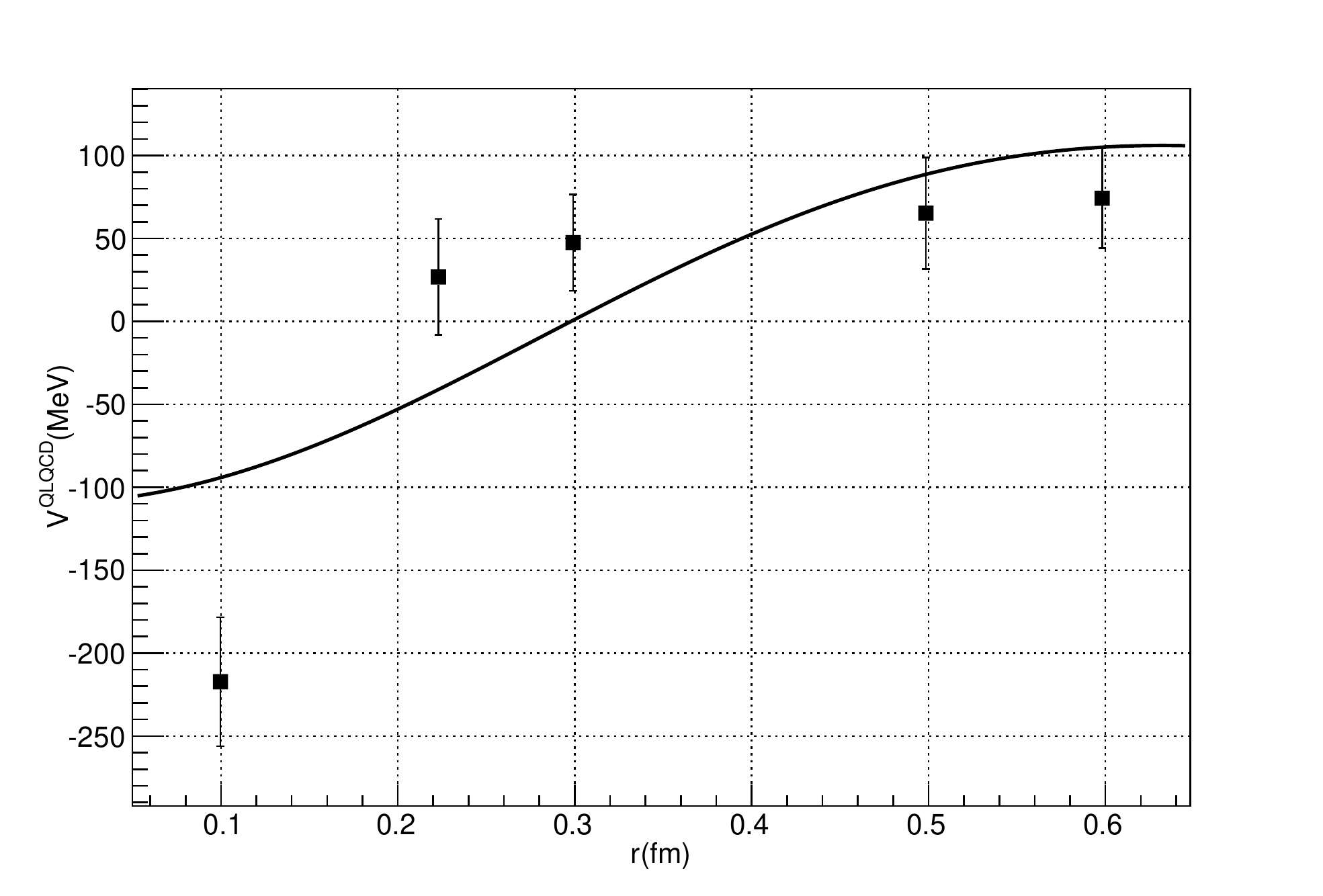} }
 }
 \caption{Fitting the $\bar B\bar B$ potentials.
 The data of the quenched lattice QCD are derived from Ref. \cite{Detmold2007} where $m_\pi=402.5\pm6.7$ MeV,
 and the error $\Delta f$ is calculated with $\sqrt{\sum_i \alpha_i^2 (\Delta g_i)^2}$ if $f=\sum_i \alpha_i g_i$.
  The curve is obtained using Eq. (\ref{eqVQQCD}) and Eq. (\ref{eqVq2r}) with parameters Eq. (\ref{eqfit}).}
 \label{figFit}
 \end{figure}

By fitting the results of the quenched lattice QCD \cite{Detmold2007}, we obtain with $\chi^2_{d.o.f}=3.7$
\begin{equation}
  a^0=0,\quad
  b^0=94\pm38,\quad
   c^0=-16\pm9,\quad
    a^1+0.16b^1=42\pm8, \quad
    c^1=-69\pm13, \quad
    \Lambda_G=708\pm(2\times10^{-9})  ~{\rm MeV}.\label{eqfit}
\end{equation}
We show the lattice data and the fitted curve in Fig. \ref{figFit}. However,
we cannot determine $a^1$ and $b^1$ respectively since the results of the lattice study are given with only one set of $m_\pi=402.5\pm6.7$ MeV.

With these values of $a^I$, $b^I$, and
$c^I$ from the lattice simulation, we can discuss the potential at
the physical pion mass now.
The potential $V_{{\bar B \bar B}^{0}}$ contains three parts up to $O(\epsilon^2)$: $V^{(2,{\rm LEC})}_{{\bar B \bar B}^{0}}$ from LECs at $O(\epsilon^2)$,
the 2$\pi$-exchange contribution $V^{(2,2\pi)}_{{\bar B \bar B}^{0}}$,
and loop corrections to the contact terms $V^{(2,{\rm cont})}_{{\bar B \bar B}^{0}}$.
Presently we cannot determine the third term since there is only one set of data with $m_\pi=402.5\pm6.7$ MeV.
We obtain the first term with the value of $b^0=94$ and $c^0=-16$, and the second term by turning off the propagation of kaon and eta.
As $|\vec q|$ goes from 0 to 300 MeV,
$V^{(2,{\rm LEC})}_{{\bar B \bar B}^{0}}$ changes from 1.8 ${\rm GeV^{-2}}$ to 0.4 ${\rm GeV^{-2}}$,
and $V^{(2,2\pi)}_{{\bar B \bar B}^{0}}$ changes from $-0.88$ ${\rm GeV^{-2}}$ to $-6.6$ ${\rm GeV^{-2}}$.
Thus the potential induced by the LECs is repulsive whereas the
2$\pi$-exchange potential is attractive. The potential induced by
the first and the second terms is repulsive but very weak at
extremely small momentum in the case without the contributions of
kaon or eta, and it becomes attractive when $|\vec q|$ is hundreds
of MeV.
If $b^0$ and $c^0$ take values in the interval $(-10,10)$,
$V^{(2,{\rm LEC})}_{{\bar B \bar B}^{0}}$ is 0.2 ${\rm GeV^{-2}}$ uppermost,
and it is smaller than $|V^{(2,2\pi)}_{{\bar B \bar B}^{0}}|$.


\end{document}